[1994/12/01]
\documentclass{aomart}
\usepackage{graphicx}
\usepackage{amsmath,amsthm, amssymb}
\usepackage[english]{babel}
\usepackage{subfigure}
\usepackage[T1]{fontenc}
\usepackage[utf8]{inputenc}
\DeclareUnicodeCharacter{0080}{~}

\chardef\bslash=`\\ 
\newcommand{\ntt}{\normalfont\ttfamily}
\newcommand{\cn}[1]{{\protect\ntt\bslash#1}}
\newcommand{\pkg}[1]{{\protect\ntt#1}}
\newcommand{\fn}[1]{{\protect\ntt#1}}
\newcommand{\env}[1]{{\protect\ntt#1}}
\newtheorem[{}\it]{thm}{Theorem}[section]

\theoremstyle{definition}

\newtheorem*[{}\it]{notation}{Notation}

\newcommand{\thmref}[1]{Theorem~\ref{#1}}
\newcommand{\secref}[1]{\S\ref{#1}}
\newcommand{\lemref}[1]{Lemma~\ref{#1}}

\newcommand{\A}{\mathcal{A}}
\newcommand{\B}{\mathcal{B}}
\newcommand{\st}{\sigma}
\newcommand{\XcY}{{(X,Y)}}
\newcommand{\SX}{{S_X}}
\newcommand{\SY}{{S_Y}}
\newcommand{\SXY}{{S_{X,Y}}}
\newcommand{\SXgYy}{{S_{X|Y}(y)}}
\newcommand{\Cw}[1]{{\hat C_#1(X|Y)}}
\newcommand{\G}{{G(X|Y)}}
\newcommand{\PY}{{P_{\mathcal{Y}}}}
\newcommand{\X}{\mathcal{X}}
\newcommand{\wt}{\widetilde}
\newcommand{\wh}{\widehat}

\DeclareMathOperator{\per}{per}

\DeclareMathOperator{\Cham}{Cham}
\DeclareMathOperator{\IM}{Im}
\DeclareMathOperator{\esssup}{ess\,sup}
\DeclareMathOperator{\meas}{meas}

\newcommand{\interval}[1]{\mathinner{#1}}

\newcommand{\eval}[2][\right]{\relax
	\ifx#1\right\relax \left.\fi#2#1\rvert}

\newcommand{\envert}[1]{\left\lvert#1\right\rvert}
\let\abs=\envert

\newcommand{\enVert}[1]{\left\lVert#1\right\rVert}
\let\norm=\enVert

\title[Dynamical behavior of Predator-Prey with Allee Effect on Both Populations and Disease in Predator]{Dynamical behavior of Predator-Prey with Allee Effect on Both Populations and Disease in Predator}
\author[Khushbu Singh]{Khushbu Singh}
\address{Department of Mathematics, NIT, Warangal-506004, India}
\email{khushbu91@student.nitw.ac.in} 
\author{K. Kaladhar}
\address{Department of Mathematics, NIT, Warangal-506004, India}
\email{kaladhar@nitw.ac.in}

\subject{primary}{matsc2020}{92D25}
\subject{primary}{matsc2020}{92D40}
\subject{primary}{matsc2020}{92D50}
\subject{primary}{matsc2020}{34D23}
\subject{primary}{matsc2020}{92D30}

\corresponding{Khushbu Singh}

\begin{document}
	
	\begin{abstract}
		In the current study, we took into account a model of nonlinear ``predator-prey'' interactions including the ``Allee effect'' on both populations and disease in the predator population. The population as a whole is split into three: the prey population, susceptible predator, and diseased predator. The ``Takagi-Sugeno (T-S) impulsive control model'' and the Fuzzy impulsive control model have been used to test the stability of the three-dimensional ``Lotka-Volterra predator-prey system'' model. Following the model's formulation, the global-stability and the fuzzy solution are examined using numerical simulations and graphical displays, together with the necessary consultation, to help comprehend the effectiveness of our suggested model.
	\end{abstract}

	\maketitle
	\tableofcontents
	
	\section{Introduction}
	
	The ``Allee effect'' and its direct effects on ecology and biomass conservation have attracted academics' attention over the past few of decades. A phenomenon known as the ``Allee Effect'' characterized by a favorable association between population size or density and an individual's fitness was first described in ecological systems by ecologist W.C. Allee in 1931 \cite{warder1931allee}. Low population density, social disorder, and elevated predation danger as a result of failing, swarming, or schooling behavior can all contribute to this effect \cite{mccarthy1997allee,dennis1989allee,wang1999competitive}. There are two different forms of ``Allee effects'', the stronger of which results in a utility of the ``Allee'' constant over which the population either increases or decreases. The weak ``Allee'' case is the other, in which the ``Allee'' constant has no threshold value. ``Allee effect'' has been investigated by numerous writers in a variety of ecological and eco-epidemiological issues \cite{stephens1999consequences,keitt2001allee,zhou2005stability,petrovskii2005regimes,morozov2006spatiotemporal,zu2010impact,zhang2018exponential,kumar2020impact}.
	
	The interaction of ecology and epidemiology has recently captured the interests of academics since infectious diseases have an impact on the ecological system, which regulates population size. There are several ``prey-predator models'' including infections. Eco-epidemic models are studied in \cite{anderson1986invasion} and \cite{hadeler1989predator} when predator populations contract the disease by consuming infected prey. The change of the ``prey-predator'' interaction in relation to diseases that affect both the prey and predator populations have been proposed and investigated by certain scholars, including Hudson \cite{hudson1992parasites}, Venturino \cite{venturino1994influence}, Xiao and Chen \cite{xiao2001analysis,xiao2001modeling}, \cite{venturino2002epidemics}, Hethcote \cite{hethcote2004predator}, Hsieh and Hsiao \cite{hsieh2008predator}, Haque and Venturino \cite{haque2009mathematical}, Zhou et al. \cite{zhou2009analysis}, Haque et al. \cite{haque2010predator,haquee2010predator,haque2011effect}, Tewa \cite{tewa2013predator}, \cite{singh2024population,singh2024populations} etc.	
	
	After Kermack-Mackendrick's pioneering work on the SIRS model in 1927, mathematical biology has increasingly turned its attention to epidemiology. Initial research in this area was based on the population of humans \cite{anderson1991infectious}. Later, ``prey-predator models'' used in epidemiological modeling in ecological systems attracted a lot of interest \cite{chattopadhyay1999predator,xiao2002ratio,chatterjee2006proper}. Wang et al. (2013) used an additive ``Allee effect'' to examine the effect of ``Allee'' on the stability of the reaction-diffusion ``prey-predator model'' without taking the infection of the predator into account\cite{wang2013allee}.

	The stability study of discrete ``predator-prey systems'' vulnerable to the ``Allee effect'', however, has not received much attention. Thus, the primary driving force behind this study is to take into account systems that an ``Allee effect'' has imposed. In this research, the ``Lotka-Volterra predator-prey model'' with an ``Allee effect'' on both populations and disease in predator is taken into consideration. We investigate the model's global-asymptotic stability using the Takagi-Sugeno model, as stated in \cite{goh1978global,tseng2001fuzzy,tong2007decentralized,zheng2009fuzzy}, then offered the graphical solutions for the problem based on considerations. Stability of the ``Lotka-Volterra predator-prey'' system with fuzzy ``impulsive control'' has only recently been studied in a small number of studies.Therefore, with the support of \cite{wang2012stability,wang2015robust,wu2015model,singh2023mathematical,singh2023stability}, a ``predator-prey'' system's stability is evaluated utilizing the ``Takagi-Sugeno'' mathematical model and fuzzy ``impulsive control''.
	
	We wish to look into the consecutive ecological concerns:
	\begin{enumerate}
		\item How do ``Allee effects'' impact the changing of both prey and predator populations?
		\item Which stability conditions allow for the survival of prey and predator?
		\item Does parameters have impact on ``prey-predator'' population?
		\item Does the ``Allee'' factor have a significant impact on the mechanism that keeps the ``predator-prey'' system stable?
	\end{enumerate}
	We will attempt to provide comprehensive answers to the questions raised above by comparing the change in the model with impulsive effects to the model without impulsive effects, as well as by getting a complete total population changing image of the proposed ``prey-predator'' interaction model with impulsive effects.
	
	\section{Model derivation}
	In this part, we've built a mathematical model of one prey population and one predator population with infection in predator that includes the ``Allee effect''.
	
	The set of ODE below can be applied to illustrate the model:
	\vspace{2mm}
	\begin{itemize}
		\item Let $x$ be the total population density of the prey.
		\item $y$ represents the entire predator population density.
		\item When an infection occurs, we assume that all predator have two classes (i) the vulnerable predator population ($y_s$) and (ii) the diseased predator population ($y_i$).
	\end{itemize}
	
	\begin{equation}\label{basic system1}
		\frac{dx}{dt}  = \frac{rx^2}{x+\beta}-\frac{ex^3}{x+\beta}-\frac{nxy_i}{a+x}-mxy_s
	\end{equation}
	\begin{equation}
		\frac{dy_s}{dt} = \lambda y_s+\frac{f_1mxy_s^2}{y_s+y_i+\beta}-By_iy_s
	\end{equation}
	\begin{equation}
		\frac{dy_i}{dt} = By_iy_s+\frac{f_2nxy_i}{a+x}-\gamma y_i
	\end{equation}
	
	where $x(0)>0$, $y_s(0)>0$, $y_i(0)>0$.
	
	The Table below lists the parameters used in the \ref{basic system1} model.
	
	\begin{table*}
		\caption{Parameters utilized in the model \ref{basic system1}}
		\footnotesize\setlength{\tabcolsep}{1pt}
		\begin{center}
			\begin{tabular}{ |c|c| }
				\hline
				Parameter & Physical significance\\
				\hline
				$r$ & Prey's intrinsic growth rate\\
				\hline
				$B$ & Disease transmission rate\\
				\hline
				$e$ & intra-species competition\\
				\hline
				$m$ & predation rate for vulnerable predator \\
				\hline
				$n$ & predation rate for infected predator \\
				\hline
				$\beta$ & Allee effect\\
				\hline
				$\lambda$ & mortality rate of vulnerable predator\\
				\hline
				$\gamma$ & mortality rate of infected predator\\
				\hline
				$f_1,f_2$ & food conversion rate\\
				\hline
				$a$ & half-saturation constant\\
				\hline
			\end{tabular}
		\end{center}
		\label{table1}
	\end{table*}
	
	In order to examine the stability of the system, the matrix differential equation is expressed as
	
	\begin{equation}\label{basic system}
		\dot{x} = Dx+\phi(x)
	\end{equation}
	where\\
	$\dot{x}$=
	$\begin{pmatrix}
		\dot{x}(t)\\ \dot{y_s}(t)\\ \dot{y_i }(t)
	\end{pmatrix}$ ,
	$D$ = $\begin{bmatrix}
		\frac{gs}{s+i+\beta} &0&0\\0&-m_1-\frac{P_2i}{s+i+\beta}&0\\0&0&-m_2
	\end{bmatrix}$,
	$\phi(x)$ =
	$\begin{bmatrix}
		-\frac{es^3}{s+i+\beta}-\frac{P_1sz}{a+s}-rsi\\rsi\\\frac{C_1sz^2}{(a+s)(z+\beta)}+\frac{C_2iz^2}{(a+i)(z+\beta)}
	\end{bmatrix}$
	
	\section{Takagi-Sugeno model}\label{sec3}
	
	\subsection{Lemma}
	``Let $\dot{x}= f(x(t))$, here the state variable is $x(t) \in R^n$, and $f \in C[R^n , R^n]$ fulfills the condition $f(0)$ = 0, is a compact vector field defined in $W \subseteq Rn$.'' Employing the methods suggested by \cite{wang2004fuzzy}, As demonstrated below, One can generate a fuzzy model for the system \ref{basic system1}:
	
	\subsubsection{``Control Rule $l$'' ($l$ = 1, 2, ...$r$):}
	``IF $z_1(t)$ is $M_{l1}$ , $z_2(t)$ is $M_{l2}$ $...$ and $z_p(t)$ is $M_{lp}$ THEN ${\dot x(t)= D_l x(t)}$, where $r$ is no. of T-S fuzzy rules, $z_1(t)$, $z_2$, ..., $z_p(t)$ are the premise variables, each $M_{lj}$ is a fuzzy set and $D_l$ $\subseteq$ $R^{n*n}$ is a constant matrix.''
	
	The following linear equation can be created by converting the non-linear equations.
	
	If $x(t)$ is $M_l$ then
	
	\begin{equation}\label{impulsive fuzzy control1}
		\dot{x}(t) = D_l x(t), t \neq \tau_j 
	\end{equation}
	\begin{equation}
		\Delta (x)= k_{lj}x(t), t=\tau_j,  l = 1,2,3...r ; j = 1,2,...
	\end{equation}
	
	where, $D_l$ = 
	$\begin{bmatrix}
		z_1 -z_2-z_3-z_4&0&0\\z_5&\lambda-z_6&0\\z_7&z_6&-\gamma
	\end{bmatrix}$
	and $z_1$, $z_2$, $z_3$, $z_4$, $z_5$, $z_6$, $z_7$ are associated with the values of $x(t)$, $y_s(t)$, $y_i(t)$ (here $\displaystyle{z_1 = \frac{rx}{x+\beta}}$, $\displaystyle{z_2 = \frac{ex^2}{x+\beta}}$, $\displaystyle{z_3 = my_s}$, $\displaystyle{z_4 = \frac{ny_i}{a+x}}$, $\displaystyle{z_5 = \frac{f_1my_s^2}{y_s+y_i+\beta}}$, $\displaystyle{z_6 = By_i}$, $\displaystyle{z_7 = \frac{f_2ny_i}{a+x}}$). $M_l$, $x(t)$, $D_l$ $\in$ $R^{3*3}$, and $r$ is the number of the IF-THEN rules, $k_{l,j}$ denotes the control of the $j^{th}$ impulsive instant, $\Delta(x)\mid_{t=\tau_j}$ = $x (\tau_j - \tau_{j-1})$)

	
	In the context of a center-average deffuzifier, the ``Takagi-Sugeno fuzzy impulsive system'' can be expressed as follows:
	\begin{equation}\label{impulsive fuzzy system1}
		\dot{x}(t) =\sum_{l=1}^{r}q_l(\zeta(t))(D_l x(t)); t \neq  \tau_j
	\end{equation}
	\begin{equation}\label{impulsive fuzzy system2}
		\Delta(x) =\sum_{l=1}^{r}q_l(\zeta(t)) k_{lj} ; t=\tau_j
	\end{equation}
	where \\
	$\displaystyle{q_l(\zeta(t)) = \omega_l(\zeta(t))/\sum_{l=1}^{r} \omega_l(\zeta(t))}$,
	and $\displaystyle{\omega_i(\zeta(t)) =  \prod_{j=1}^{p} M_{ij}(\zeta(t)), l=1,2,...,r}$\\
	here, $\displaystyle{q_l(\zeta(t)) \geq 0}$, $\displaystyle{\sum_{l=1}^{r}q_l (\zeta(t)) = 1, l = 1,2,...,r}$
	
	\section{Numerical Simulation}
	
	Analytical research cannot be finished without numerical confirmation of the results. This section \eqref{basic system} presents computer simulations of the system's solutions. These numerical solutions validate our analytical results and are also important from a practical perspective. Because most biological systems are intricate, they ought to be modeled using a descriptive, fuzzy logical approach. Using the suggested impulsive ``Takagi-Sugeno model'', ``predator-prey systems'' with functional responses and impulsive effects are investigated. This section discusses the general stability of the analyzed intra-species competition ``predator-prey model'' \eqref{basic system}. Due to the complexity of biological processes, non-linear, and unpredictable, it is best to explain them utilizing a fuzzy logical approach and linguistic description.

	The membership functions \cite{wang2004fuzzy} were produced as follows using the fuzzy impulsive ``Takagi-Sugeno'' design model on the \eqref{basic system}:
	
	$\displaystyle{M_1 = \frac{z_1}{(\frac{r*d_1}{d_1+\beta})}}$, $\displaystyle{M_2=\frac{(\frac{r*d_1}{d_1+\beta})-z_1}{(\frac{r*d_1}{d_1+\beta})}}$,$\displaystyle{N_1 =\frac{z_2}{(\frac{e*d_1^2}{d_1+\beta})}}$, $\displaystyle{N_2=\frac{(\frac{e*d_1^2}{d_1+\beta})-z_2}{(\frac{e*d_1^2}{d_1+\beta})}}$,
	
	$\displaystyle{K_1 = \frac{z_3}{(m*d_2)}}$, $\displaystyle{K_2 =\frac{{(m*d_2)}-z_3}{(m*d_2)}}$, $\displaystyle{L_1 = \frac{z_4}{(\frac{n*d_3}{d_1+a})}}$, $\displaystyle{L_2 = \frac{(\frac{n*d_3}{d_1+a})-z_4}{(\frac{n*d_3}{d_1+a})}}$, 
	
	$\displaystyle{Q_1 = \frac{z_5}{(\frac{f_1*m*d_2^2}{d_2+d_3+\beta})}}$, $\displaystyle{Q_2=\frac{(\frac{f_1*m*d_2^2}{d_2+d_3+\beta})-z_5}{(\frac{f_1*m*d_2^2}{d_2+d_3+\beta})}}$, $\displaystyle{R_1 = \frac{z_6}{B*d_3}}$, $\displaystyle{R_2=\frac{B*d_3-z_6}{B*d3}}$,
	
	$\displaystyle{S_1=\frac{z_7}{(\frac{f_2*n*d_3}{d_1+a})}}$, $\displaystyle{S_2=\frac{(\frac{f_2*n*d_3^2}{(d_1+a)})-z_7}{(\frac{f_2*n*d_3^2}{(d_1+a)})}}$\\
	
	Hence, Deffuzification is given as
	\begin{equation}
		\dot x(t) =  \sum_{l=1}^{r}q_l(\zeta(t))(D_l x(t))
	\end{equation}
	here $q_l's$ are given as $\displaystyle{q_l(\zeta(t)) = \omega_l(\zeta(t)) / \sum_{l=1}^{r} \omega_l(\zeta(t))}$, and $\displaystyle{\omega_l(\zeta(t)) =  \prod_{j=1}^{p} M_{lj}(\zeta(t))}$, $l$ = 1 to 127, $j$ = 1 to 7, where $M_{lj}'s$ are membership functions.
	
	An appropriate representation of the non-linear system in the region [0,10]x[0,10]x[0,10] is provided by this fuzzy model.
	
	\begin{equation}
		\frac{dx}{dt} =\frac{rs^2}{x+\beta}-\frac{ex^3}{x+\beta}-\frac{nxy_i}{a+x}-mxy_s
	\end{equation}
	\begin{equation}
		\frac{dy_s}{dt} = \lambda y_s+\frac{f_1mxy_s^2}{y_s+y_i+\beta}-By_iy_s
	\end{equation}
	\begin{equation}
		\frac{dy_i}{dt} = By_iy_s+\frac{f_2nxy_i}{a+x}-\gamma y_i
	\end{equation}
	
	To compute it, values for the parameters are taken at $r=0.5$, $e=0.5$, $\lambda=1.5$, $\beta=1.0$, $m=0.25$, $n=0.25$, $d_1=10$, $d_2=10$, $d_3=10$, $a=0.2$, $f_1=0.2$, $f_2=0.1$, $\gamma=0.1$, $B=0.2$, in \ref{impulsive fuzzy control1} to obtain the eigen-values of $[D_l^T + D_l] (l=1,2,3...r)$ described in (\cite{singh2023mathematical}). Clearly, $max(\lambda_i$)=$\lambda(\alpha)=2.54$, then, we've taken $diag [-0.99,-0.99]$ as ``impulsive control'' matrix in a manner that $\beta_j=||I+K||=0.01$. 
	It has been confirmed that the \ref{impulsive fuzzy control1} system exists and is globally-stable at $\epsilon$=1.5, $\delta_j$=0.1 (at the previously given values, $\displaystyle{ln(\epsilon\beta_j)+\lambda(\alpha)\delta_j=-1.643 < 0}$). Assuming that the parameters of the system are $r=1.5$, $\lambda=0.5$, $\beta=2.0$, $e=0.5$, $m=1.5$, $n=0.5$, $d_1=10$, $d_2=10$, $d_3=10$, $a=2$, $f_1=0.8$, $f_2=0.5$, $\gamma=2.5$, $B=2.5$, since $max(\lambda_i$) =$\lambda(\alpha)=59.25$, $\implies$$ln(\epsilon\beta_j)+\lambda(\alpha)\delta_j = 4.028 > 0$ for $\beta_j=0.01$, $\epsilon$=1.5, $\delta_j$=0.1. For the above specified parameter values, it is seen that the ``prey-predator model'' is unstable.
	
	\begin{table*}
		\caption{Stability of the system in various circumstances}
		\footnotesize\setlength{\tabcolsep}{1pt}
		\begin{center}
			\begin{tabular}{ |c|c|c|c|c|c|c|c|c|c|c|c|c|c|c|c|c| }
				\hline
				$r$ & $e$ & $\beta$ & $\lambda$ & $\gamma$ & $f_1$ & $f_2$ & $m$ & $n$ & $B$ & $a$ & $d_1$ & $d_2$ & $d_3$ & max($\lambda_i$)=$\lambda(\alpha)$ & $ln(\epsilon\beta)+\lambda(\alpha)\delta_j$ & conclusion\\
				\hline
				0.5 & 0.5 & 1.0 & 1.5 & 0.1 & 0.2 & 0.1 & 0.25 & 0.25 & 0.2 & 0.2 & 10.0 & 10.0 & 10.0 & 2.54 & -1.643 & stable\\
				\hline
				1.5 & 0.5 & 2.0 & 1.5 & 0.5 & 0.3 & 0.1 & 0.5 & 0.25 & 0.5 & 2.0 & 10.0 & 10.0 & 10.0 & 9.44 & -0.953 & stable \\
				\hline
				0.5 & 2.5 & 1.0 & 0.5 & 0.5 & 0.8 & 0.5 & 2.5 & 1.5 & 0.5 & 1.0 & 10.0 & 10.0 & 1.0 & 16.498 & -0.2472 & stable\\
				\hline
				1.5 & 0.5 & 2.0 & 0.5 & 2.5 & 0.8 & 0.5 & 1.5 & 0.5 & 2.5 & 2.0 & 10.0 & 10.0 & 10.0 & 59.25 & 4.028 & unstable\\
				\hline
			\end{tabular}
		\end{center}
		\label{stability_table}
	\end{table*}

	\section{Findings}
	
\begin{figure}
		\centering
		{\includegraphics[height=0.35\textheight, width=0.49\textwidth]{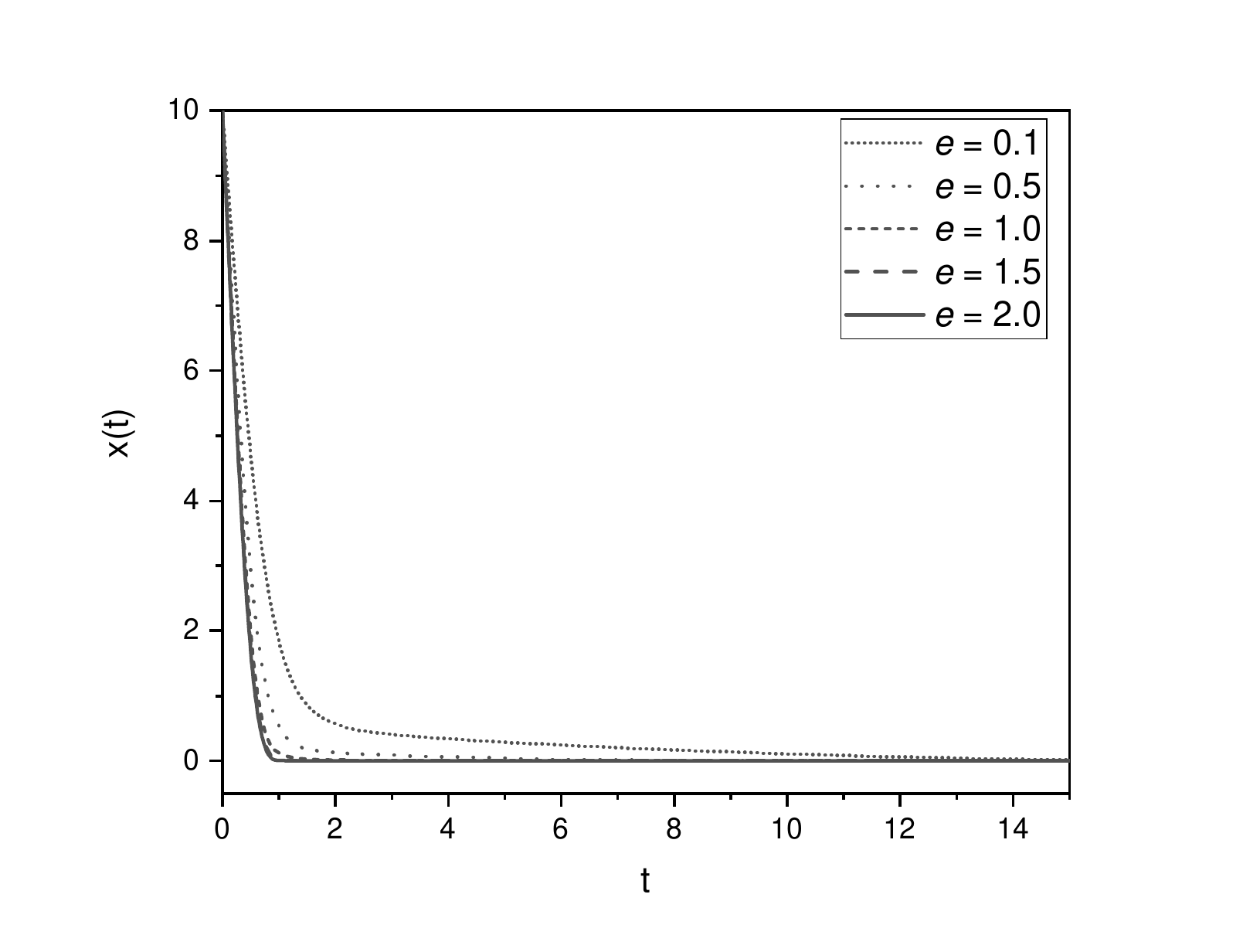}}\\
		{\includegraphics[height=0.35\textheight, width=0.49\textwidth]{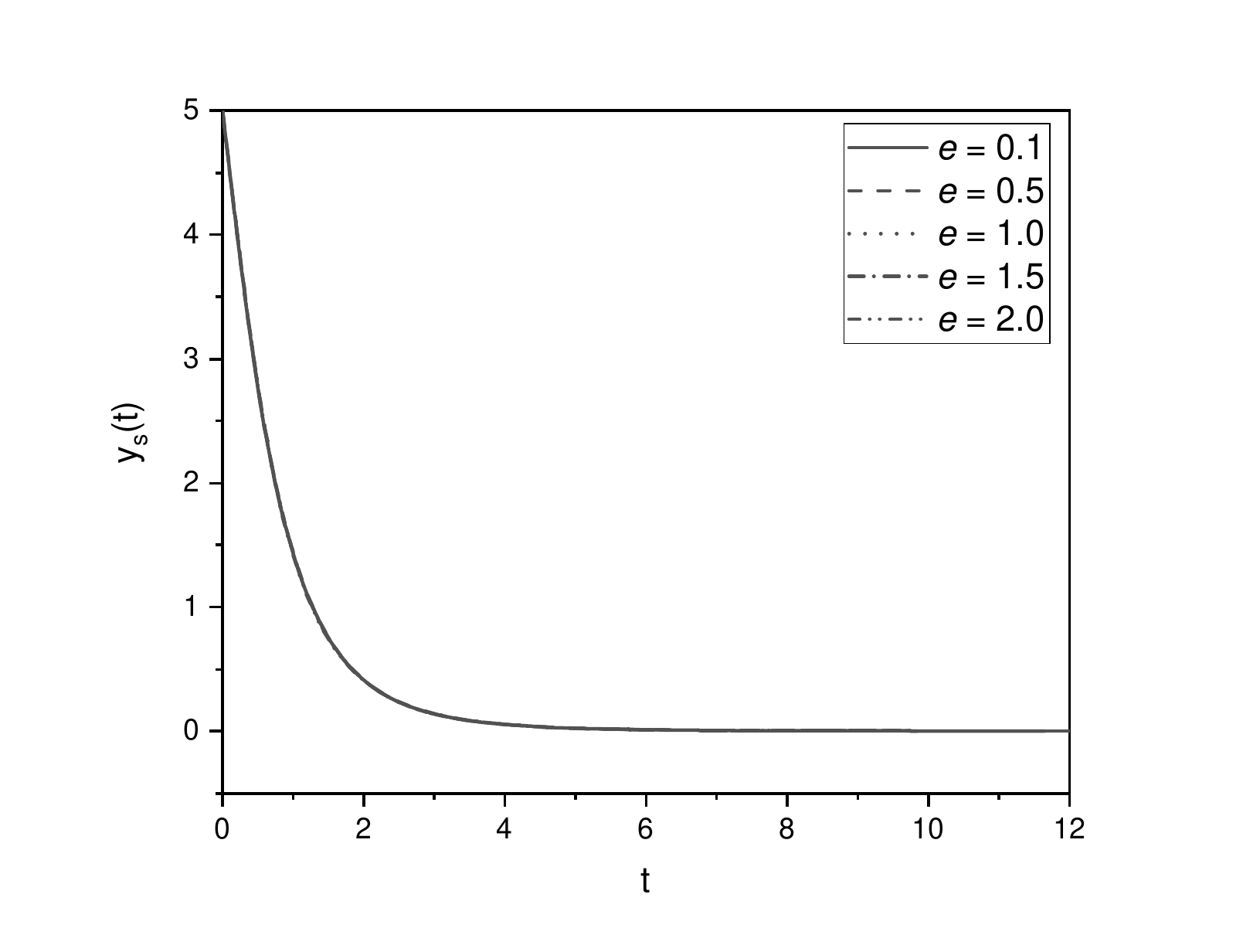}}
		{\includegraphics[height=0.35\textheight, width=0.49\textwidth]{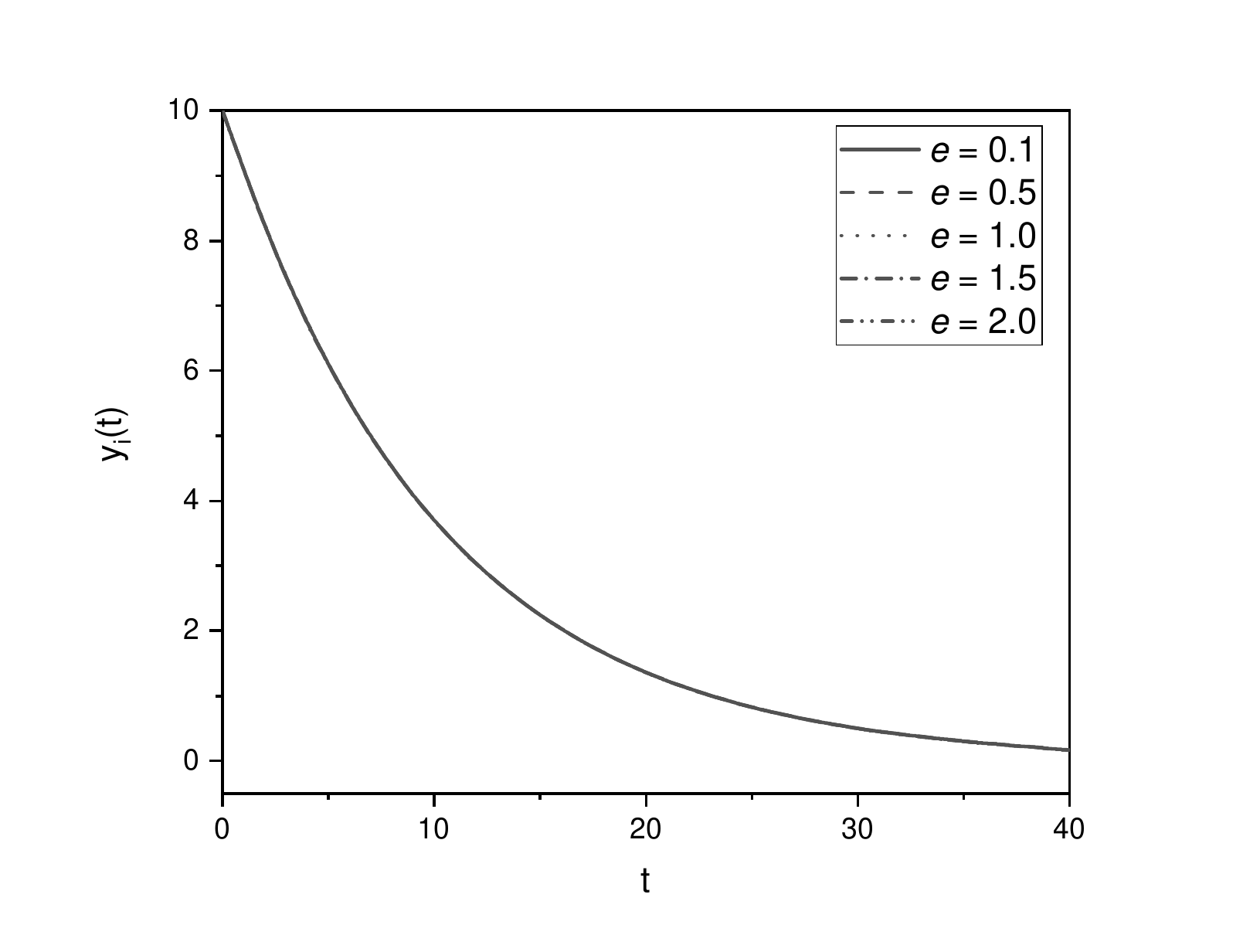}}
		\caption{$e$ effect on ``prey-predator system'' under ``impulsive control''}
		\label{figure 1}
	\end{figure}

\begin{figure}
		\centering
		{\includegraphics[height=0.35\textheight, width=0.49\textwidth]{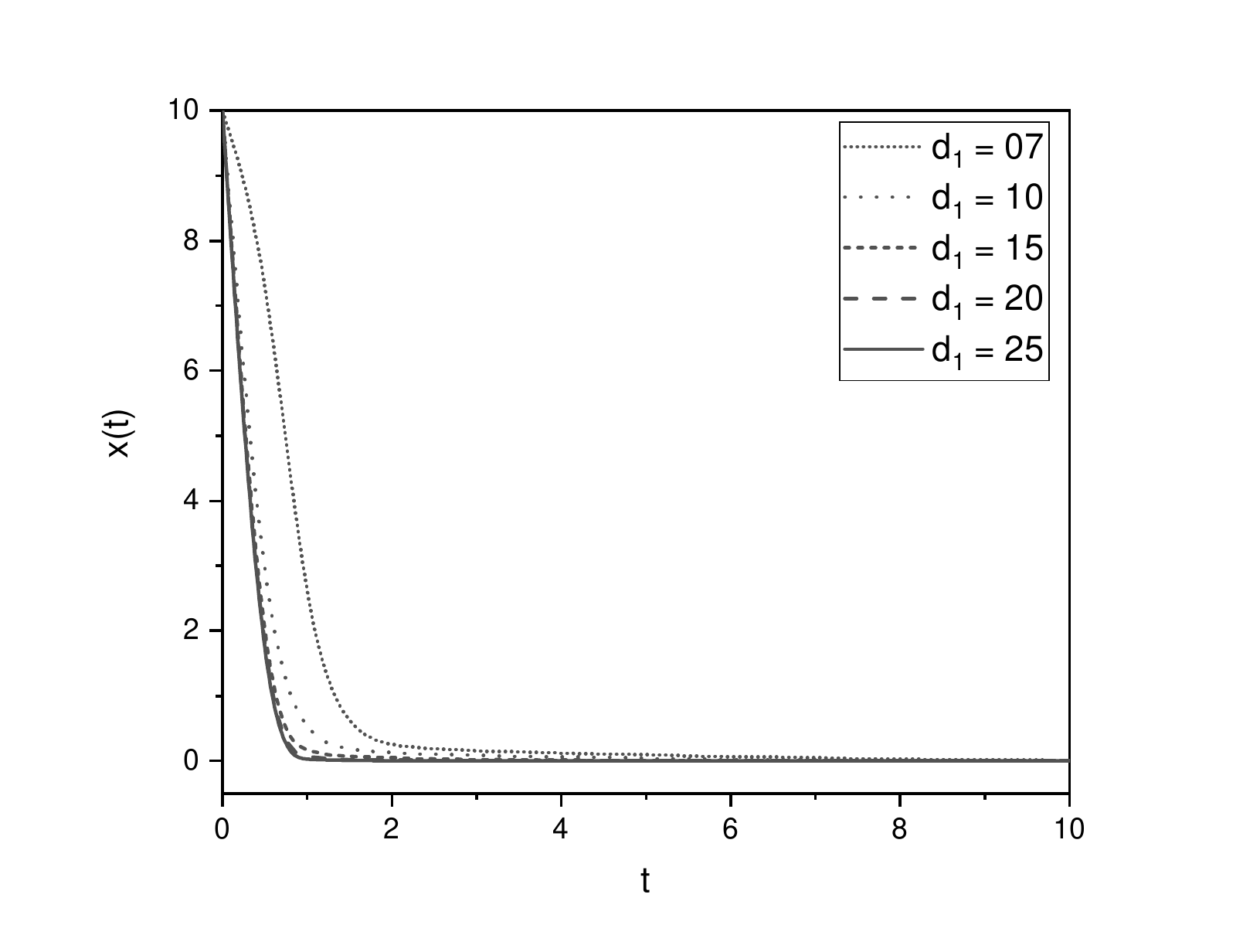}}\\
		{\includegraphics[height=0.35\textheight, width=0.49\textwidth]{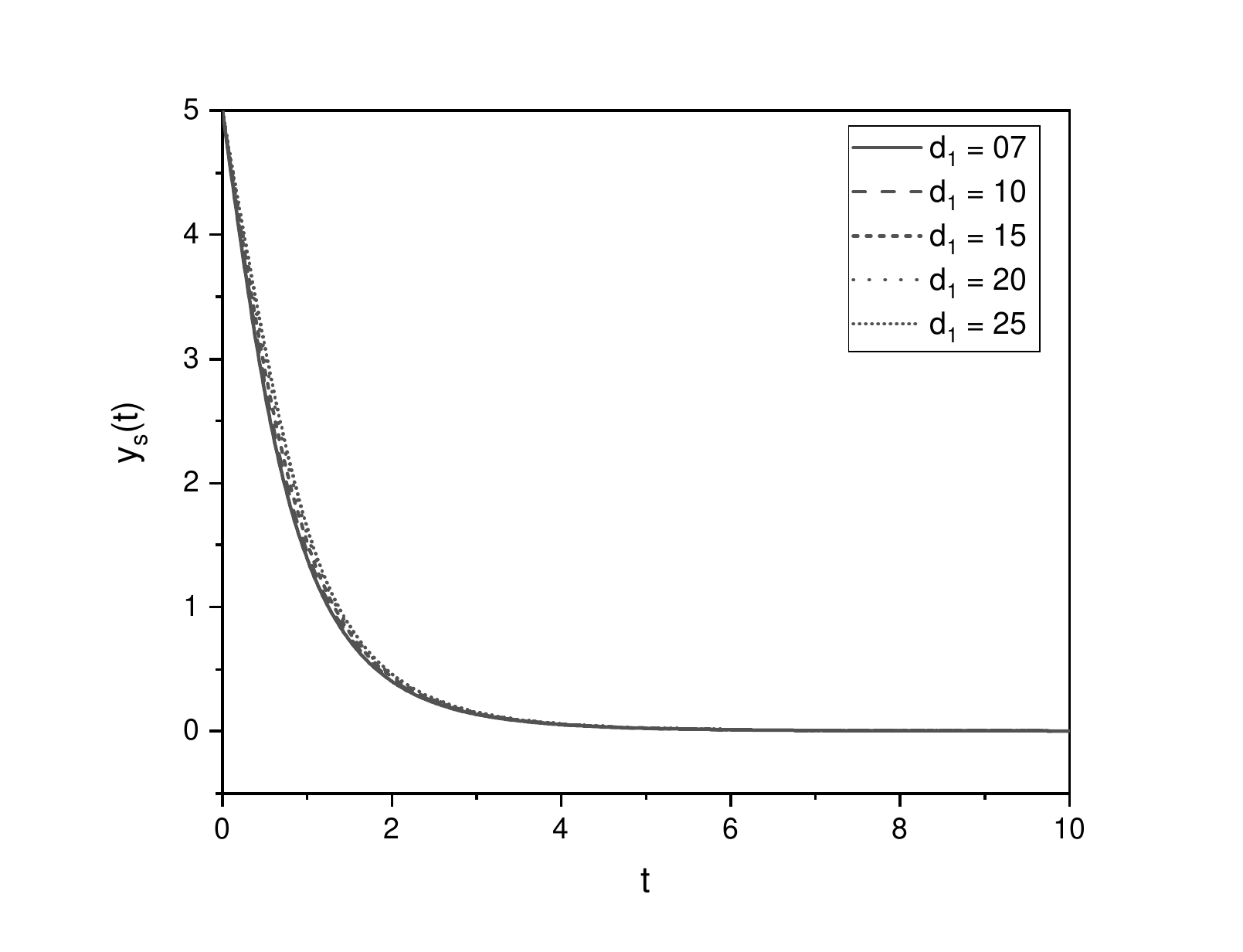}}
		{\includegraphics[height=0.35\textheight, width=0.49\textwidth]{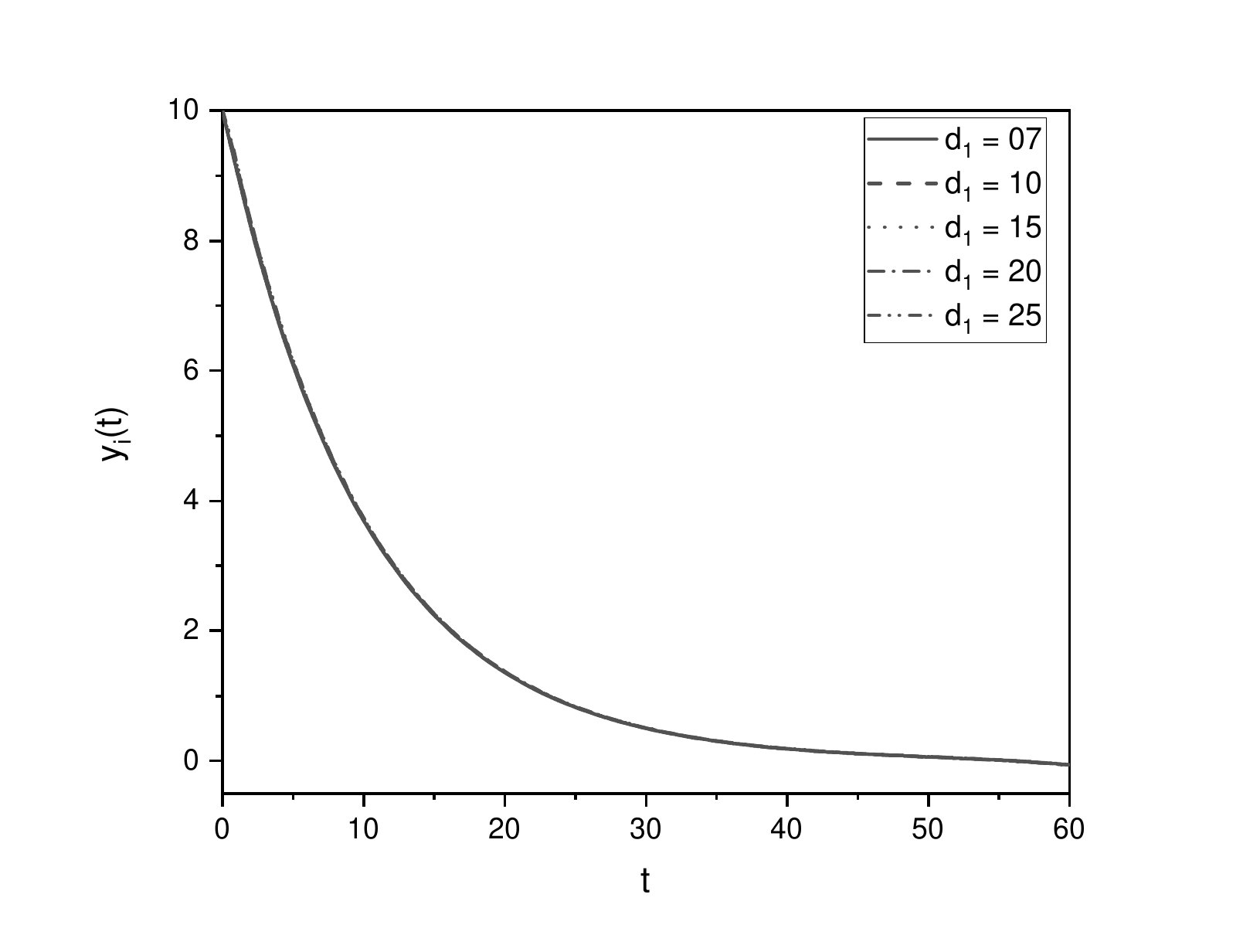}}
		\caption{$d_1$ effect on ``prey-predator system'' under ``impulsive control''.}
		\label{figure 2}
	\end{figure}

\begin{figure}
		\centering
		{\includegraphics[height=0.35\textheight, width=0.49\textwidth]{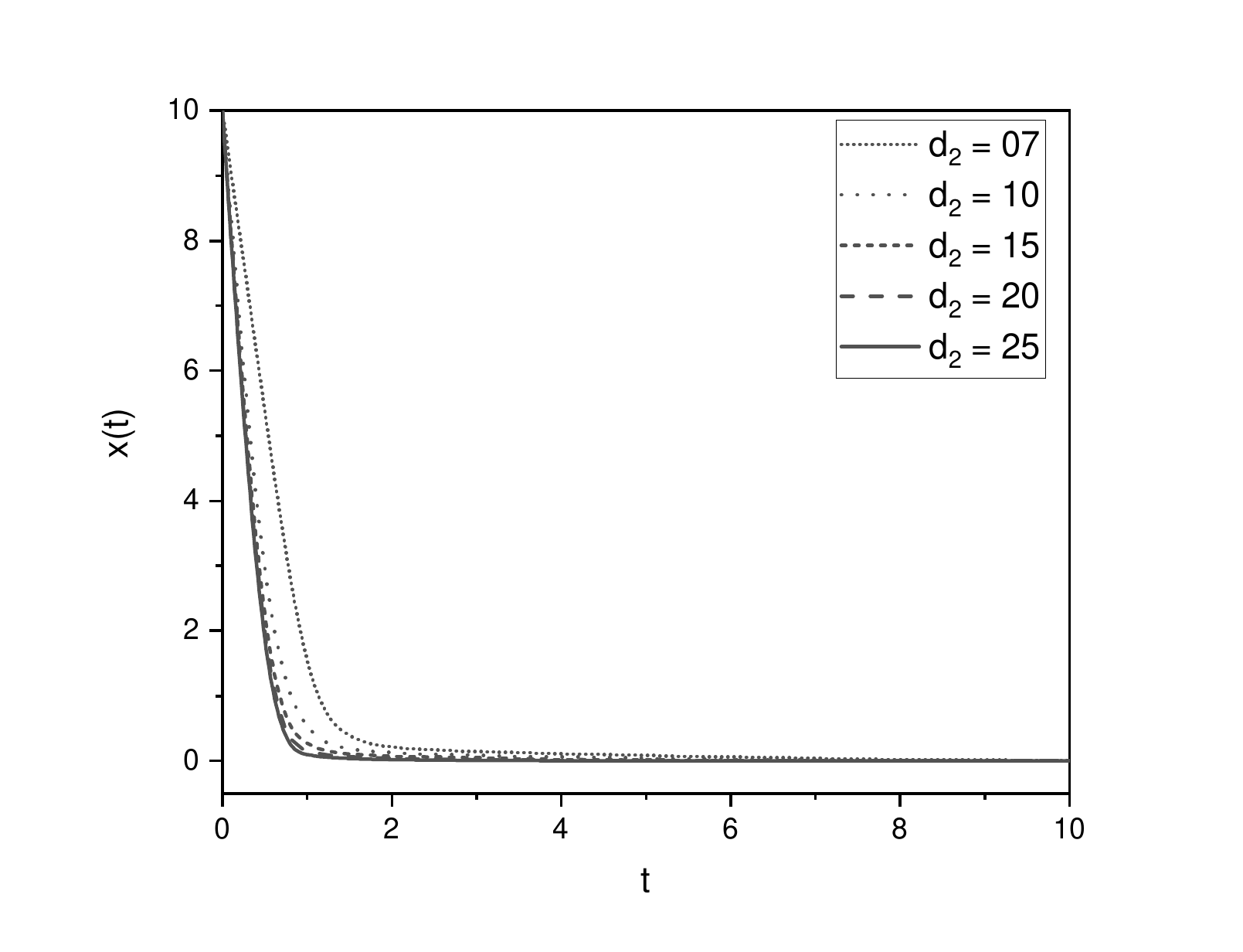}}\\
		{\includegraphics[height=0.35\textheight, width=0.49\textwidth]{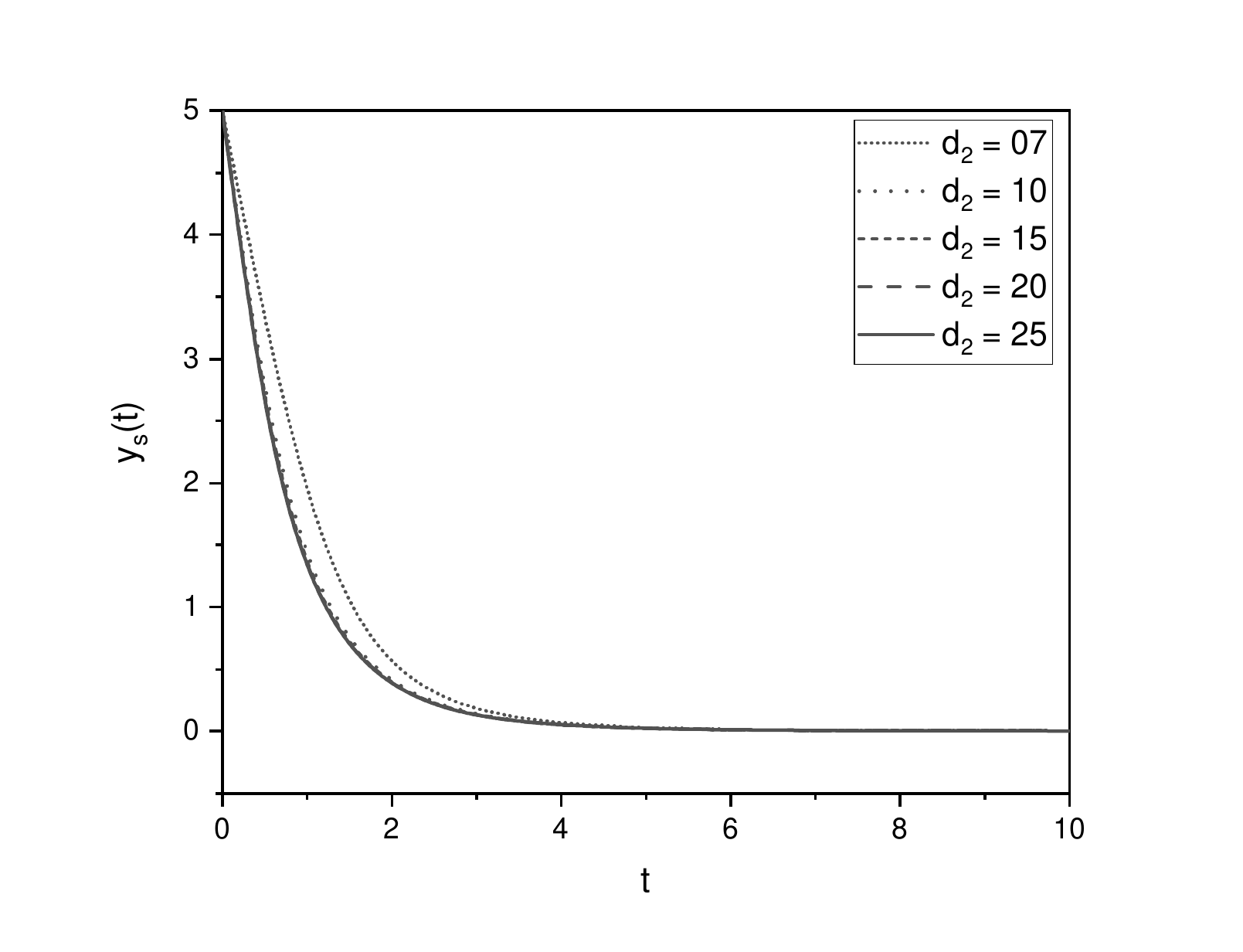}}
		{\includegraphics[height=0.35\textheight, width=0.49\textwidth]{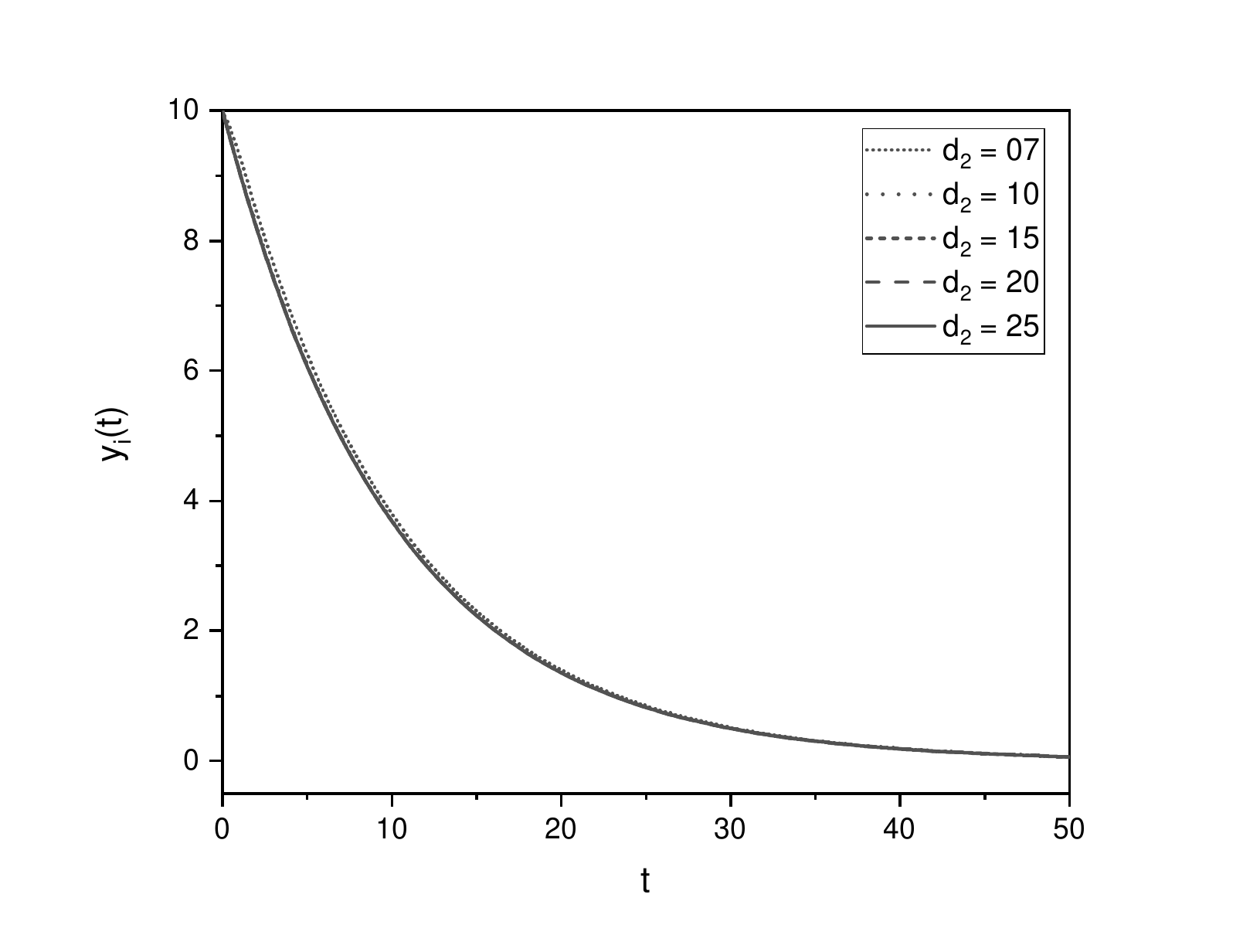}}
		\caption{$d_2$ effect on ``prey-predator system'' under ``impulsive control''.}
		\label{figure 3}
	\end{figure}

\begin{figure}
		\centering
		{\includegraphics[height=0.35\textheight, width=0.49\textwidth]{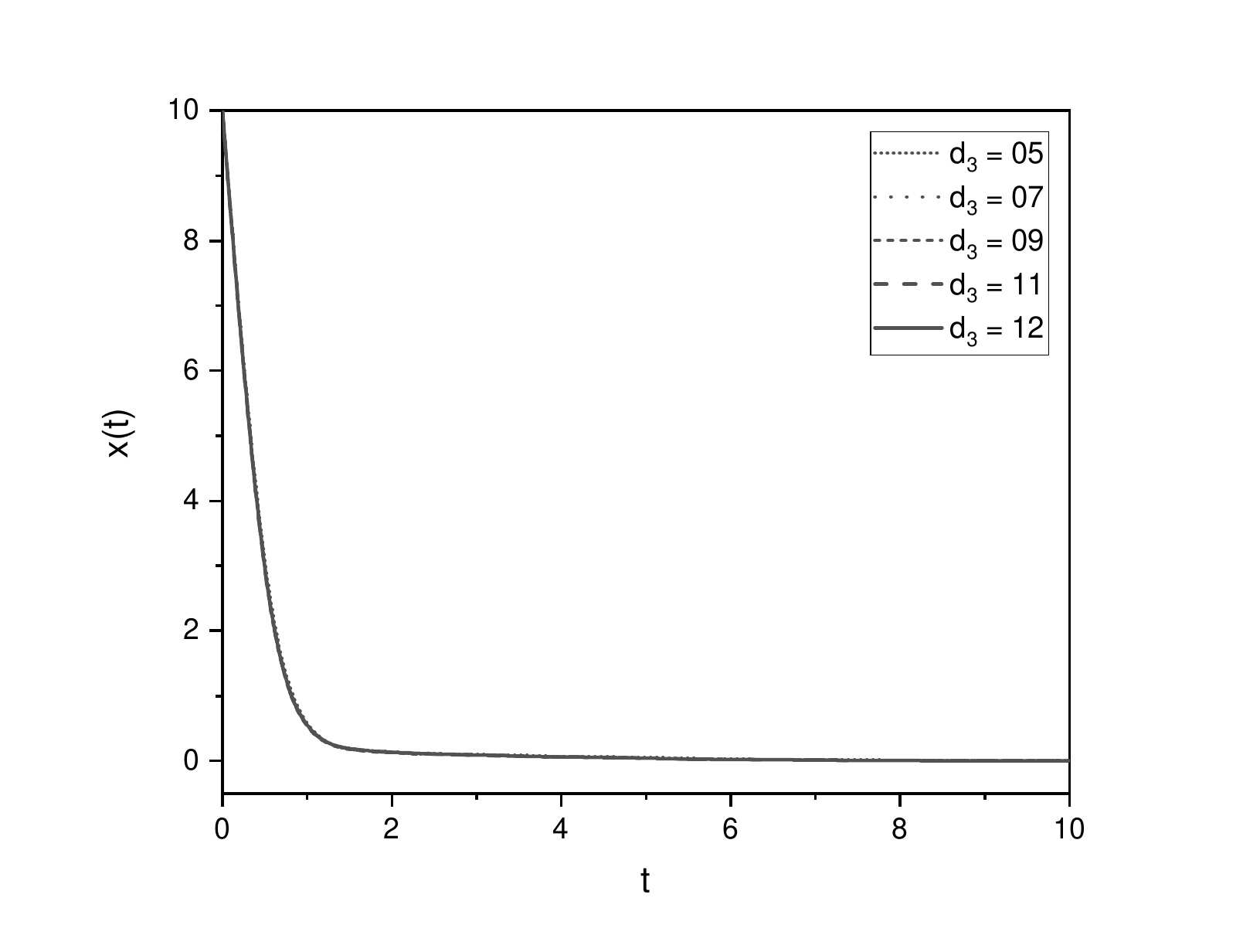}}\\
		{\includegraphics[height=0.35\textheight, width=0.49\textwidth]{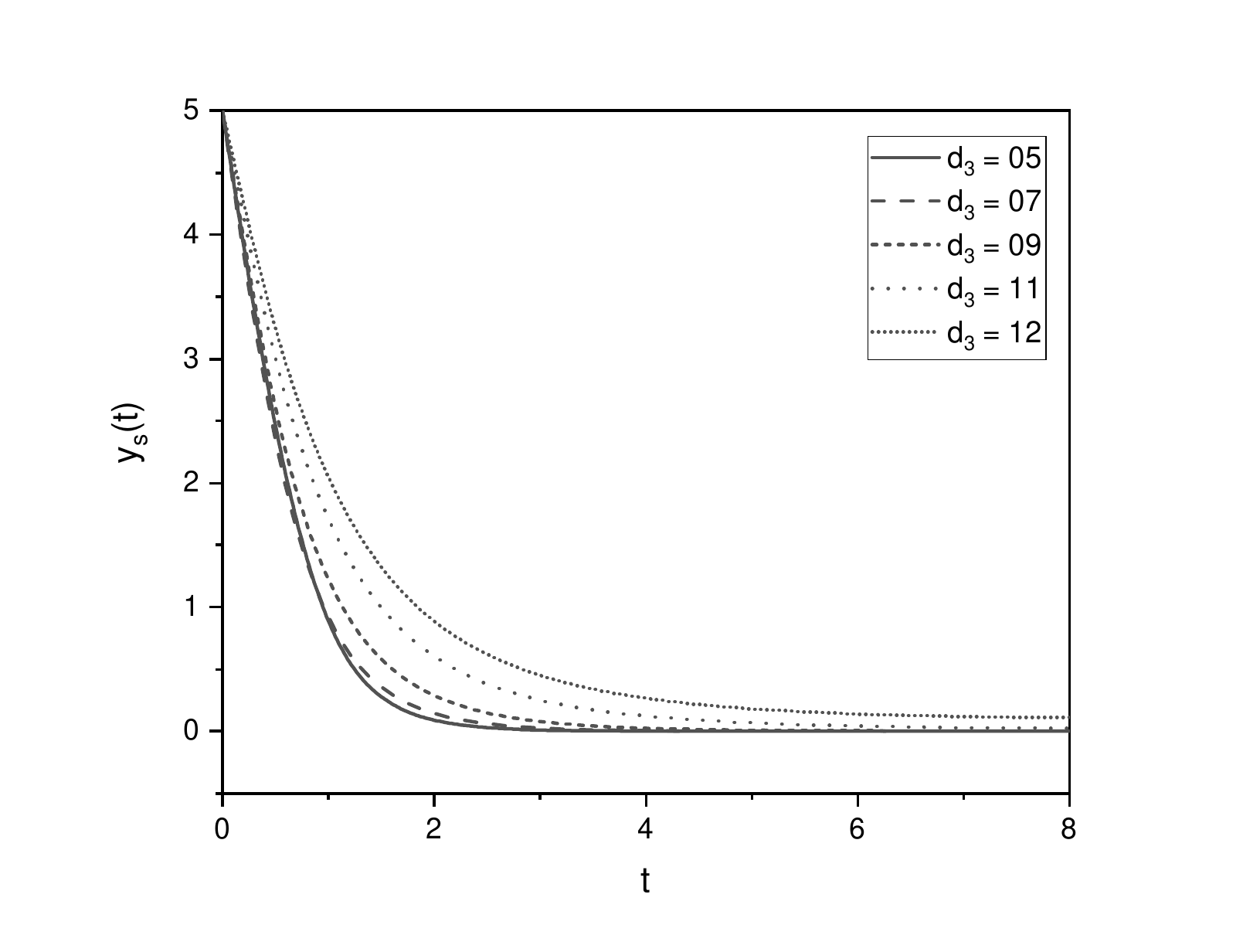}}
		{\includegraphics[height=0.35\textheight, width=0.49\textwidth]{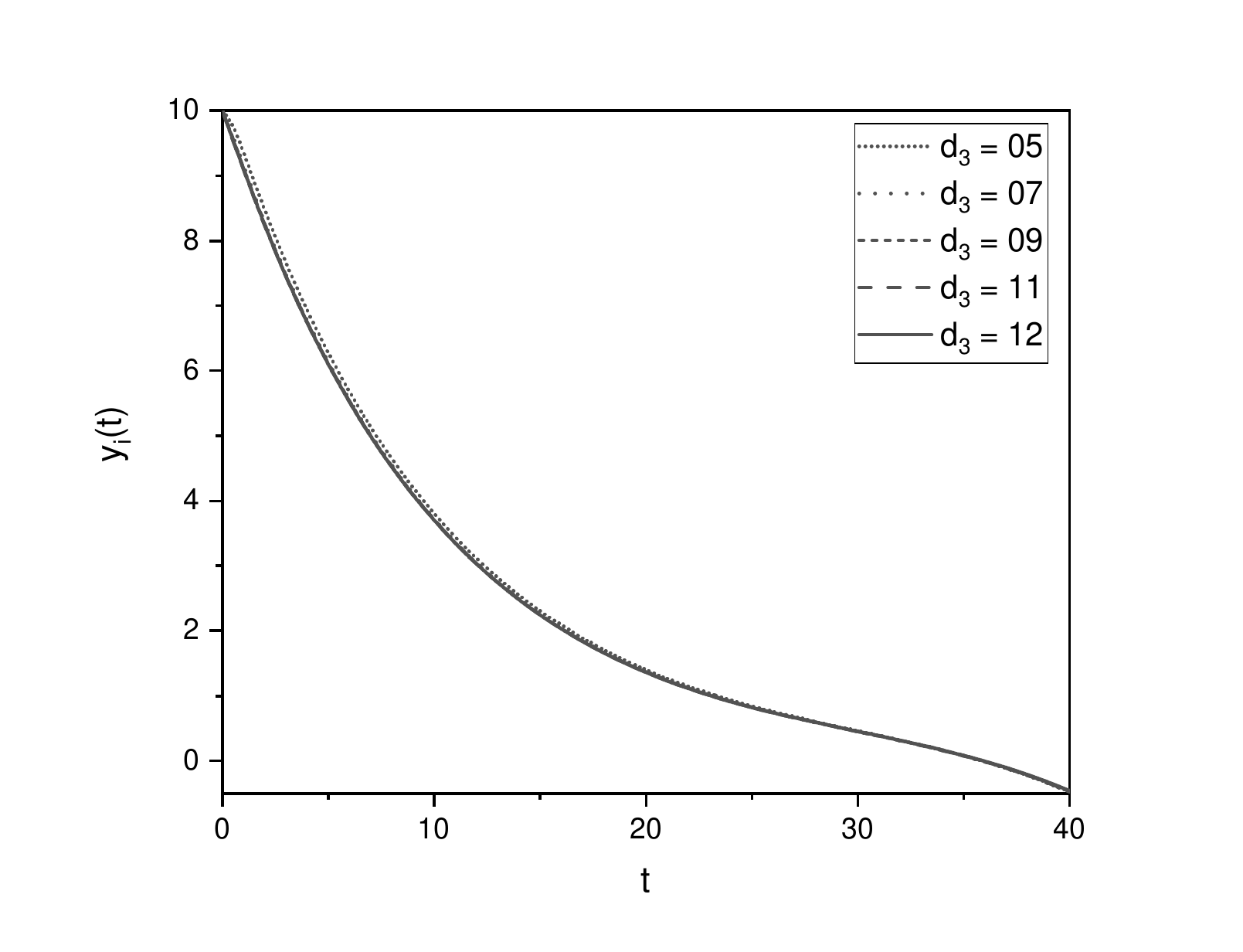}}
		\caption{$d_3$ effect on ``prey-predator system'' under ``impulsive control''.}
		\label{figure 4}
	\end{figure}

\begin{figure}
		\centering
		{\includegraphics[height=0.35\textheight, width=0.49\textwidth]{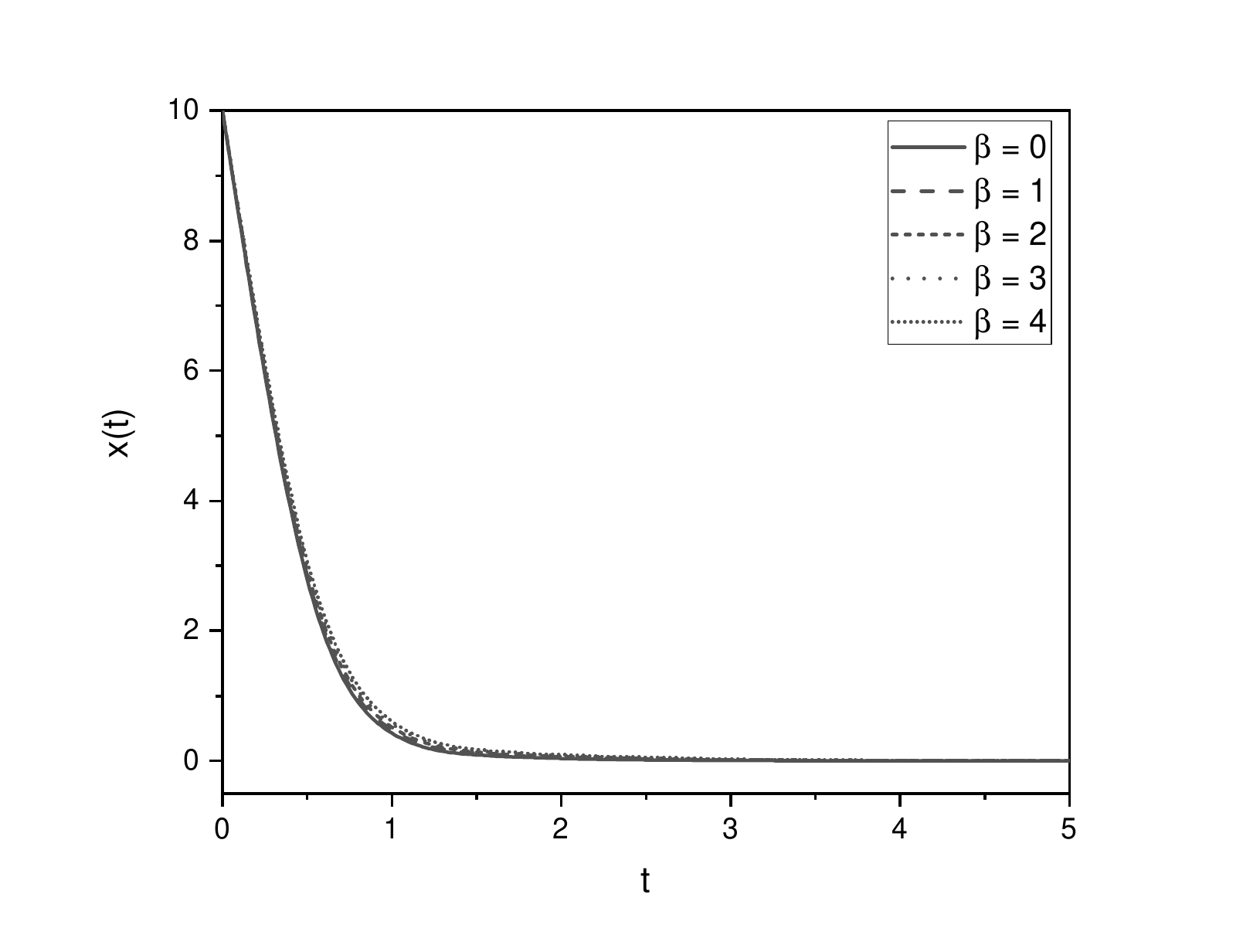}}\\
		{\includegraphics[height=0.35\textheight, width=0.49\textwidth]{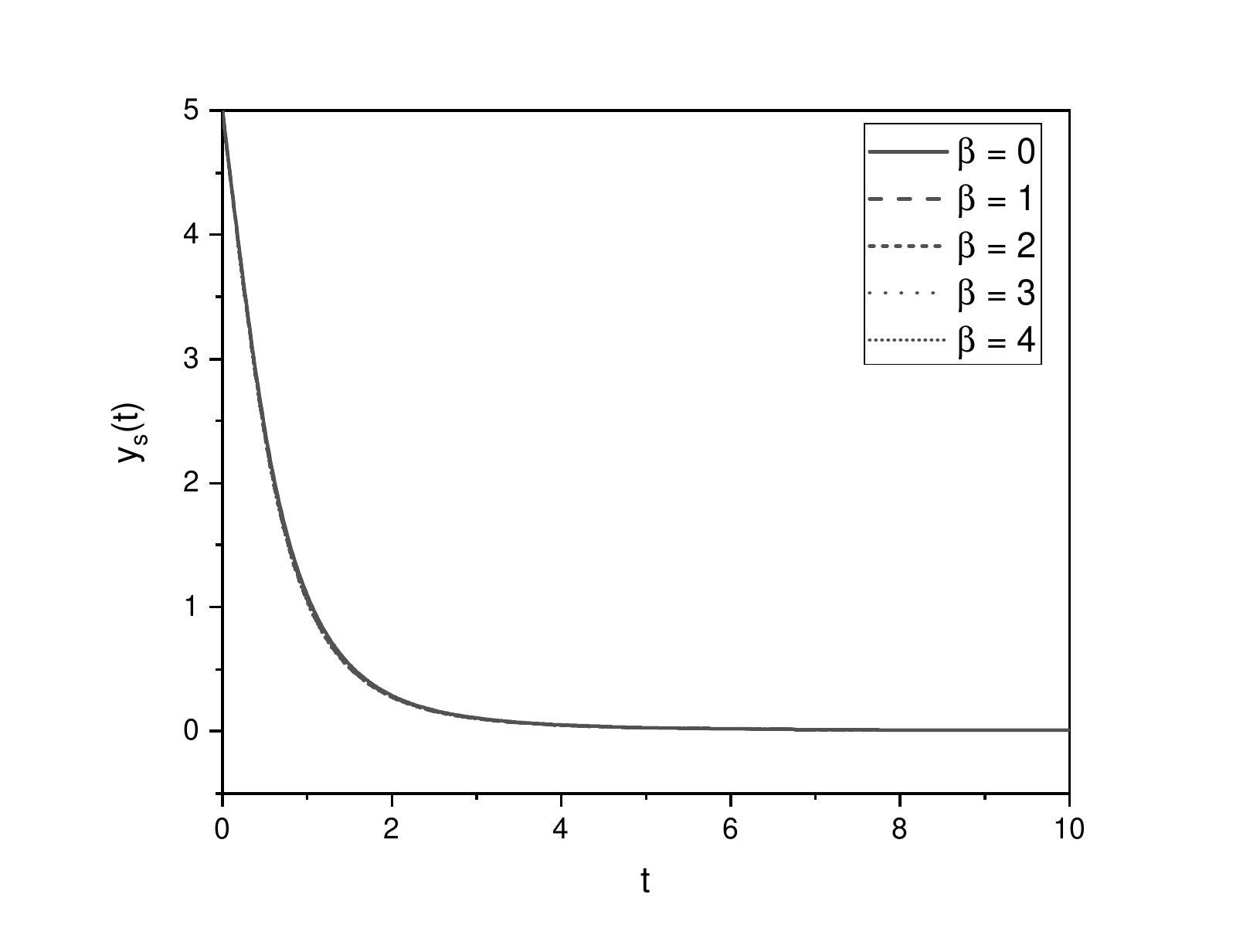}}
		{\includegraphics[height=0.35\textheight, width=0.49\textwidth]{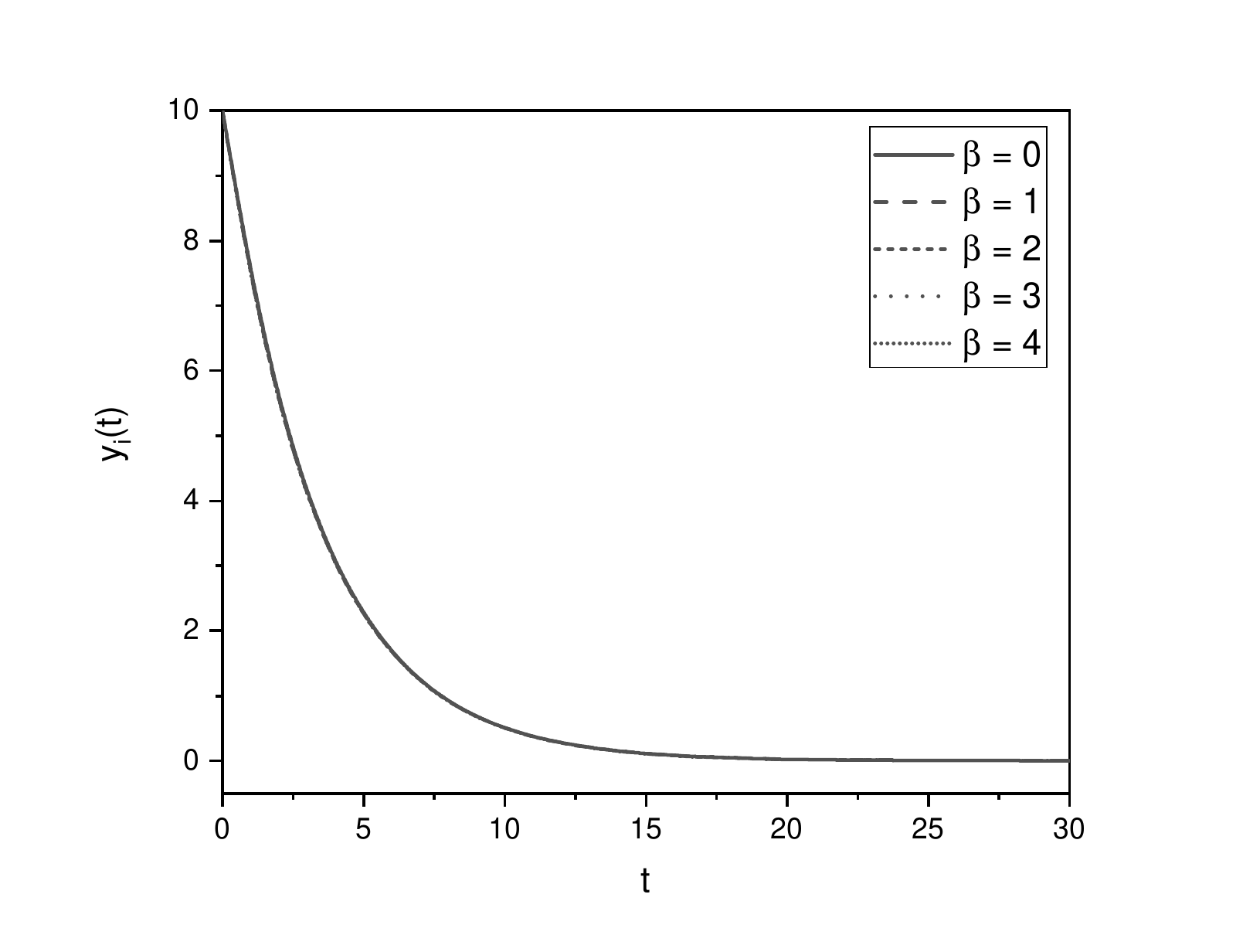}}
		\caption{$\beta$ effect on ``prey-predator system'' under ``impulsive control''.}
		\label{figure 5}
	\end{figure}

\begin{figure}
		\centering
		{\includegraphics[height=0.35\textheight, width=0.49\textwidth]{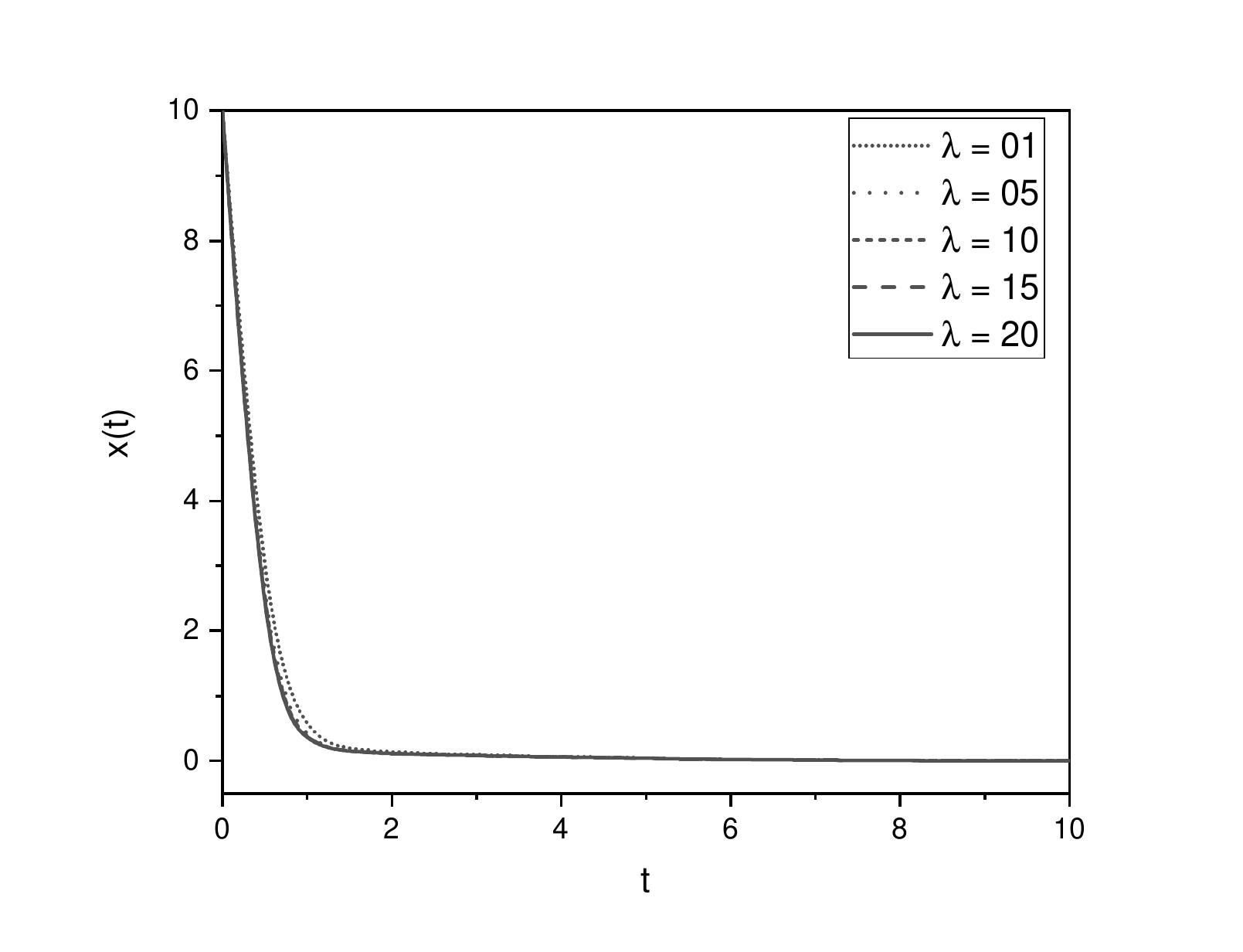}}\\
		{\includegraphics[height=0.35\textheight, width=0.49\textwidth]{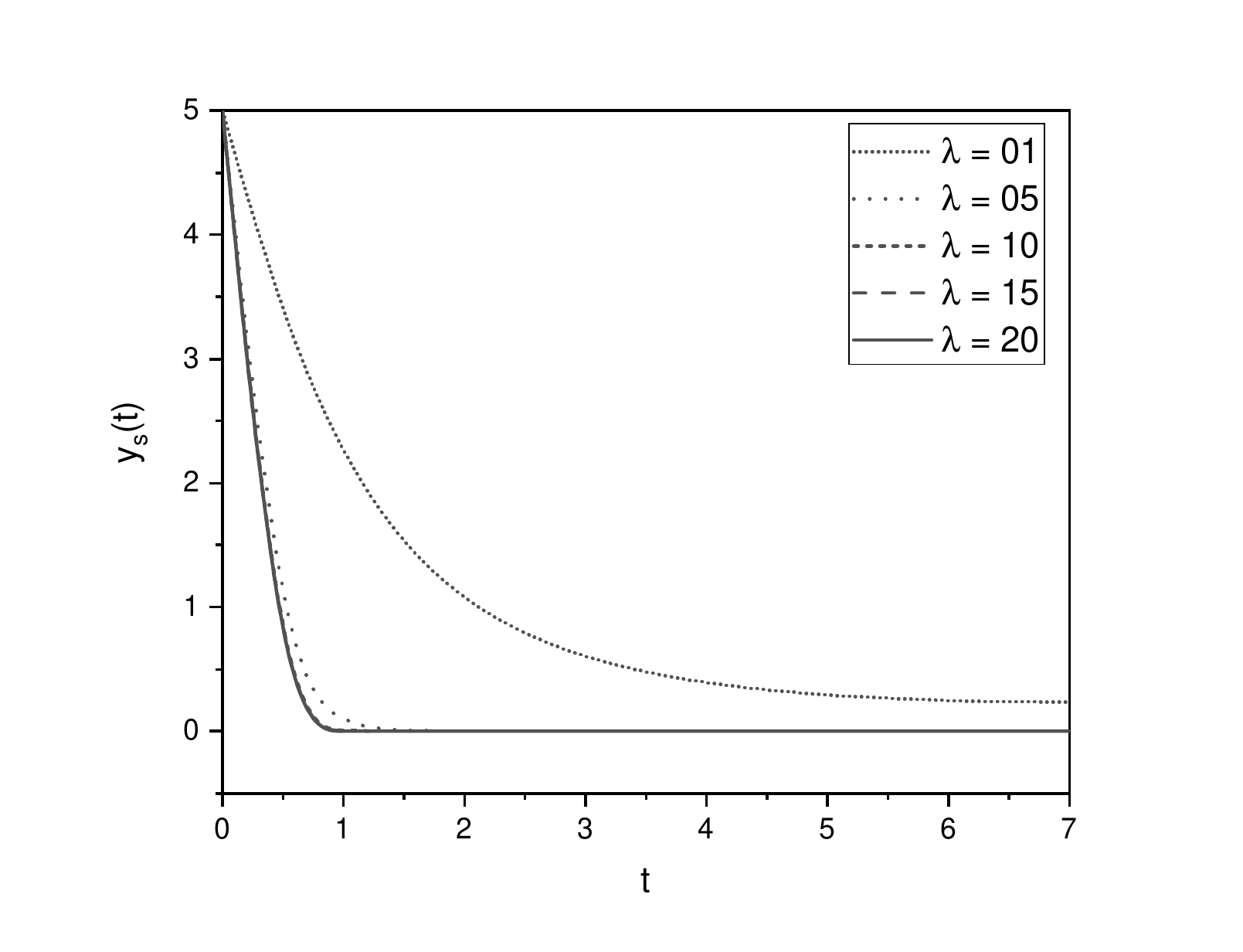}}
		{\includegraphics[height=0.35\textheight, width=0.49\textwidth]{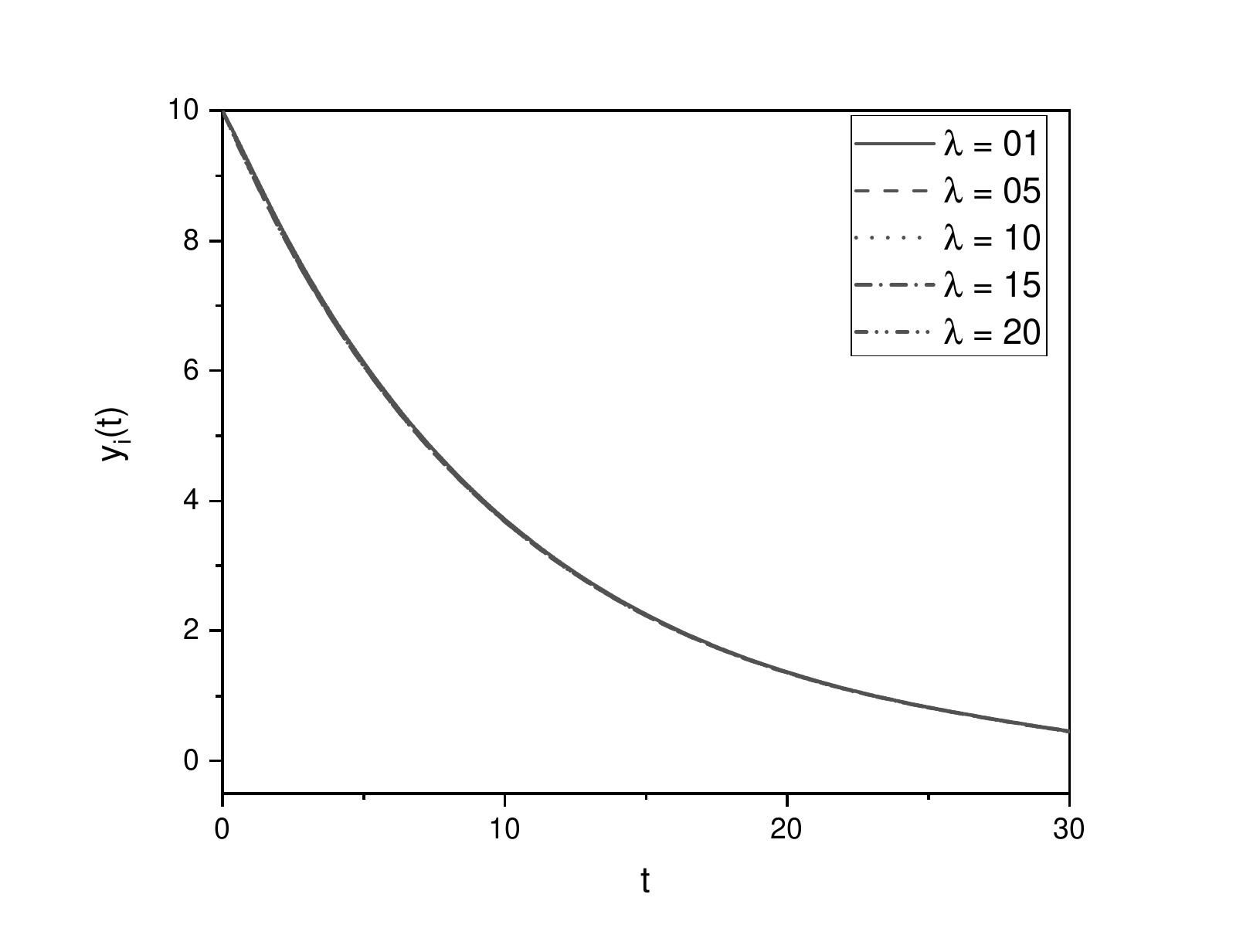}}
		\caption{$\lambda$ effect on ``prey-predator system'' under ``impulsive control''.}
		\label{figure 6}
	\end{figure}

\begin{figure}
		\centering
		{\includegraphics[height=0.35\textheight, width=0.49\textwidth]{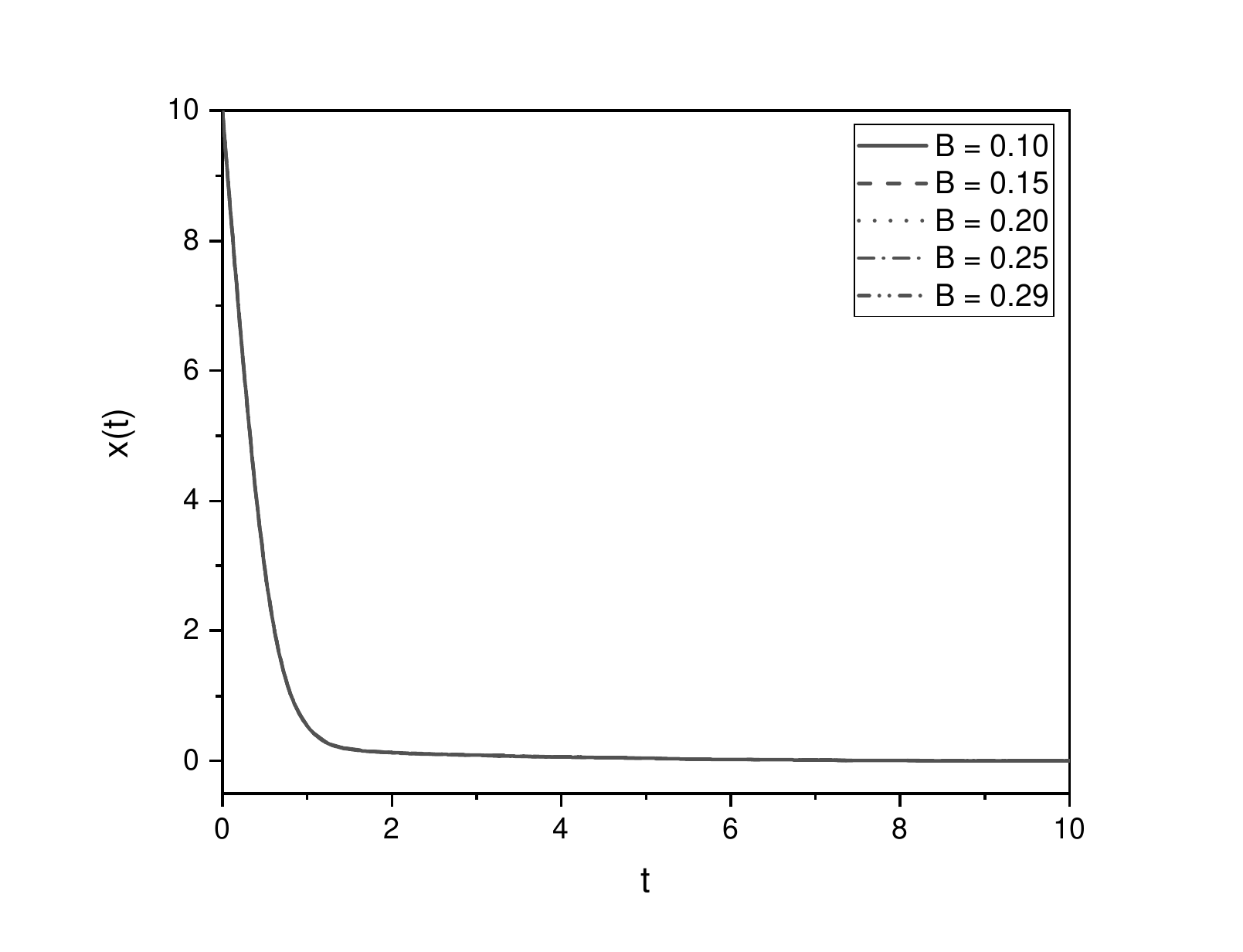}}\\
		{\includegraphics[height=0.35\textheight, width=0.49\textwidth]{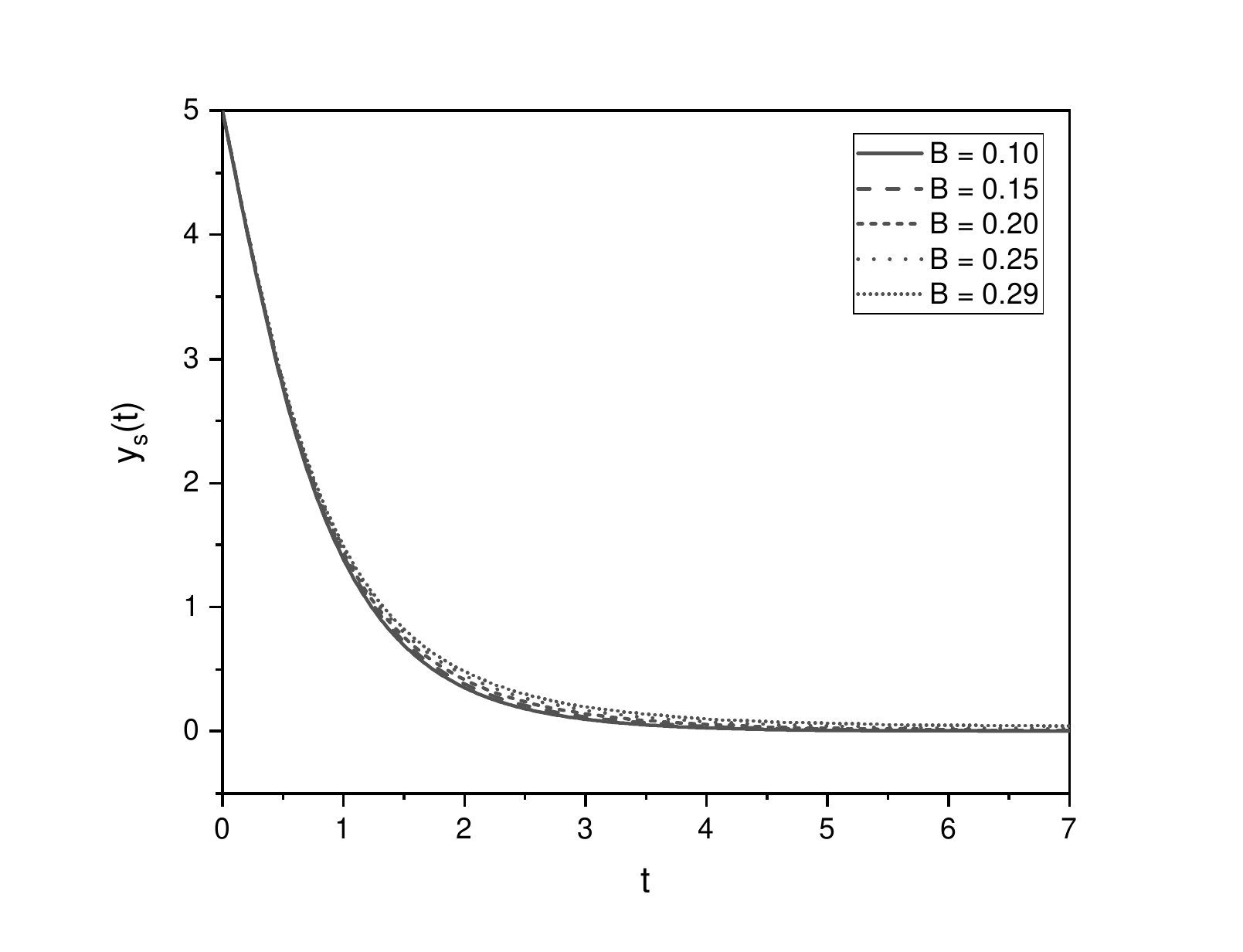}}
		{\includegraphics[height=0.35\textheight, width=0.49\textwidth]{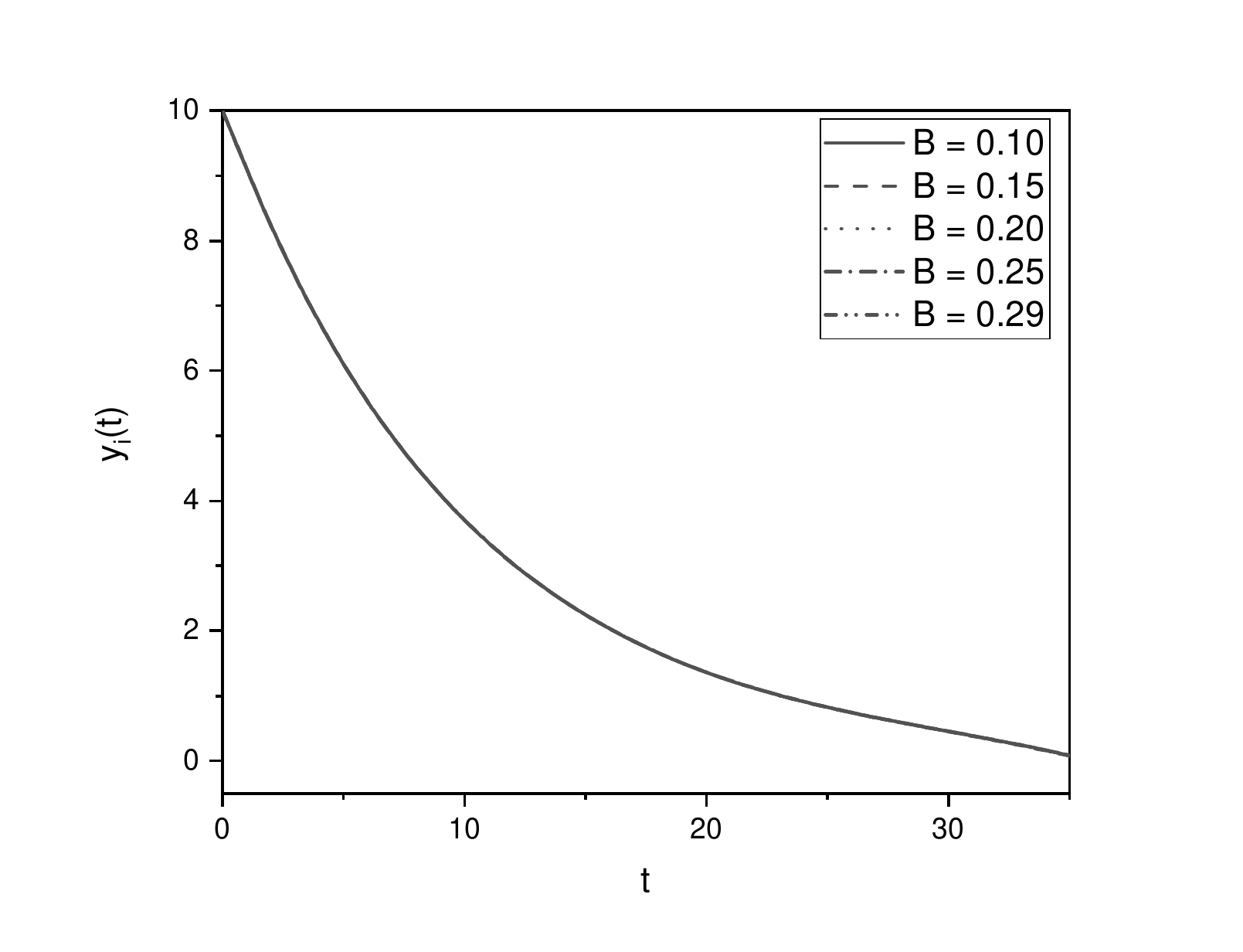}}
		\caption{$B$ effect on ``prey-predator system'' under ``impulsive control''.}
		\label{figure 7}
	\end{figure}

\begin{figure}
		\centering
		{\includegraphics[height=0.35\textheight, width=0.49\textwidth]{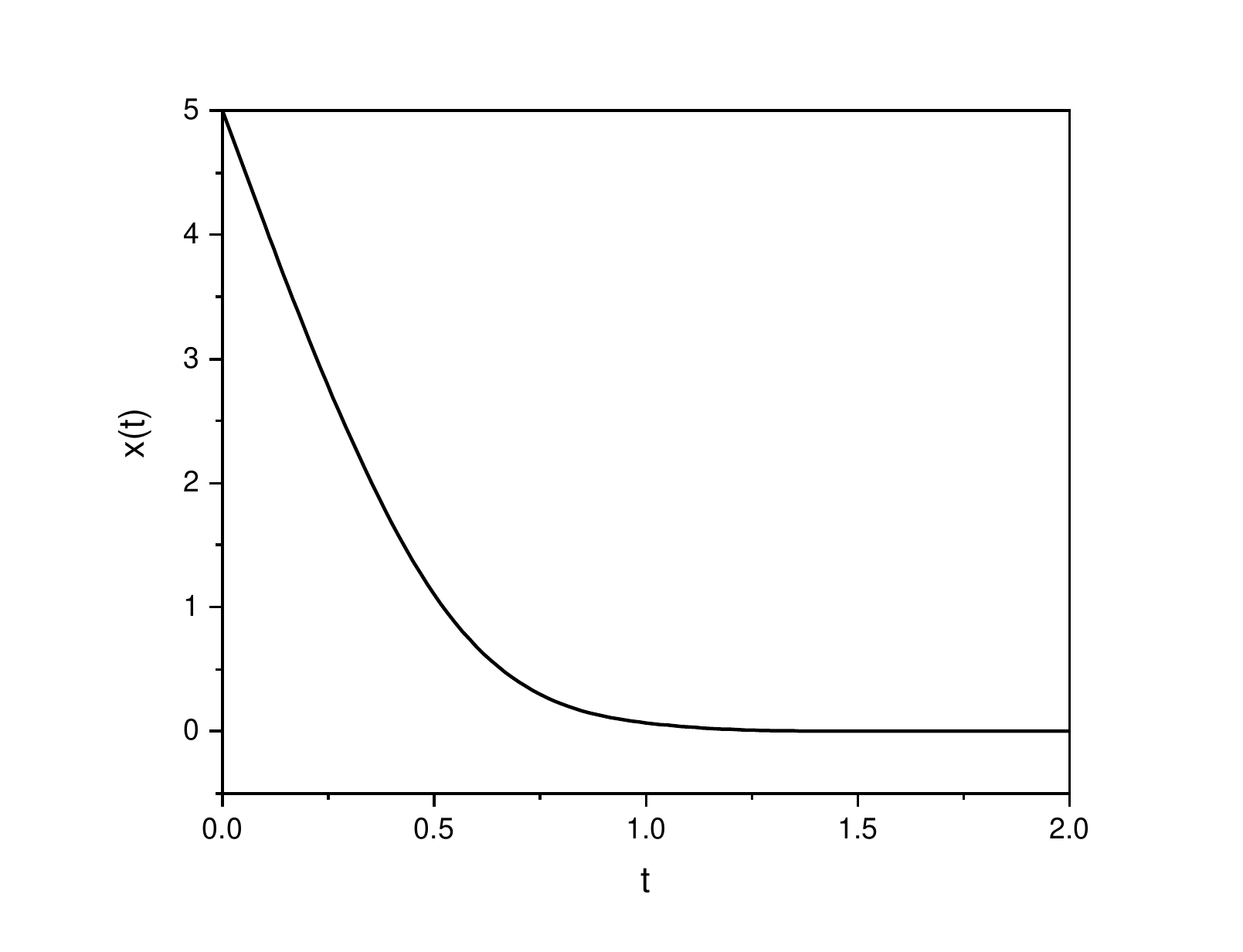}}\\
		{\includegraphics[height=0.35\textheight, width=0.49\textwidth]{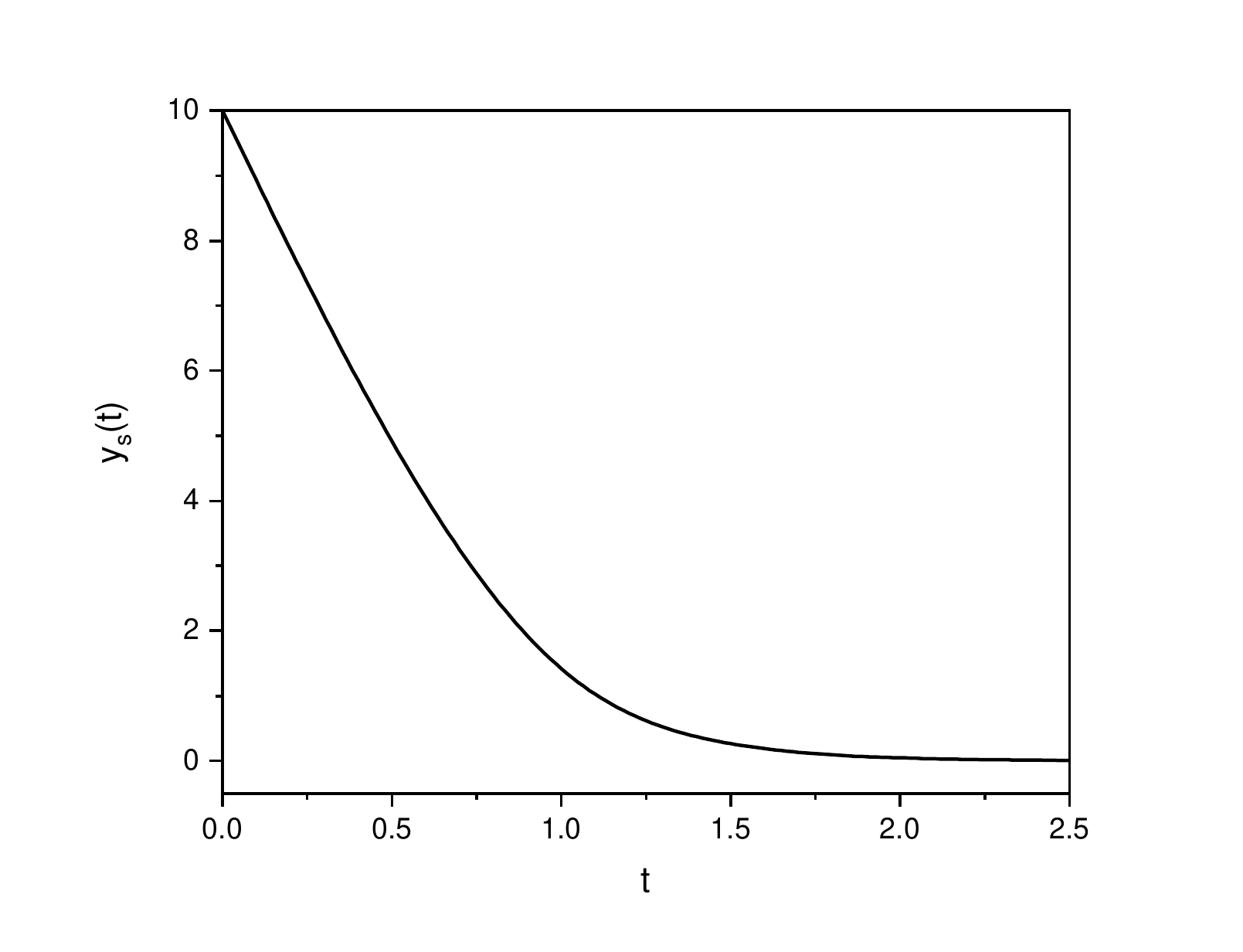}}
		{\includegraphics[height=0.35\textheight, width=0.49\textwidth]{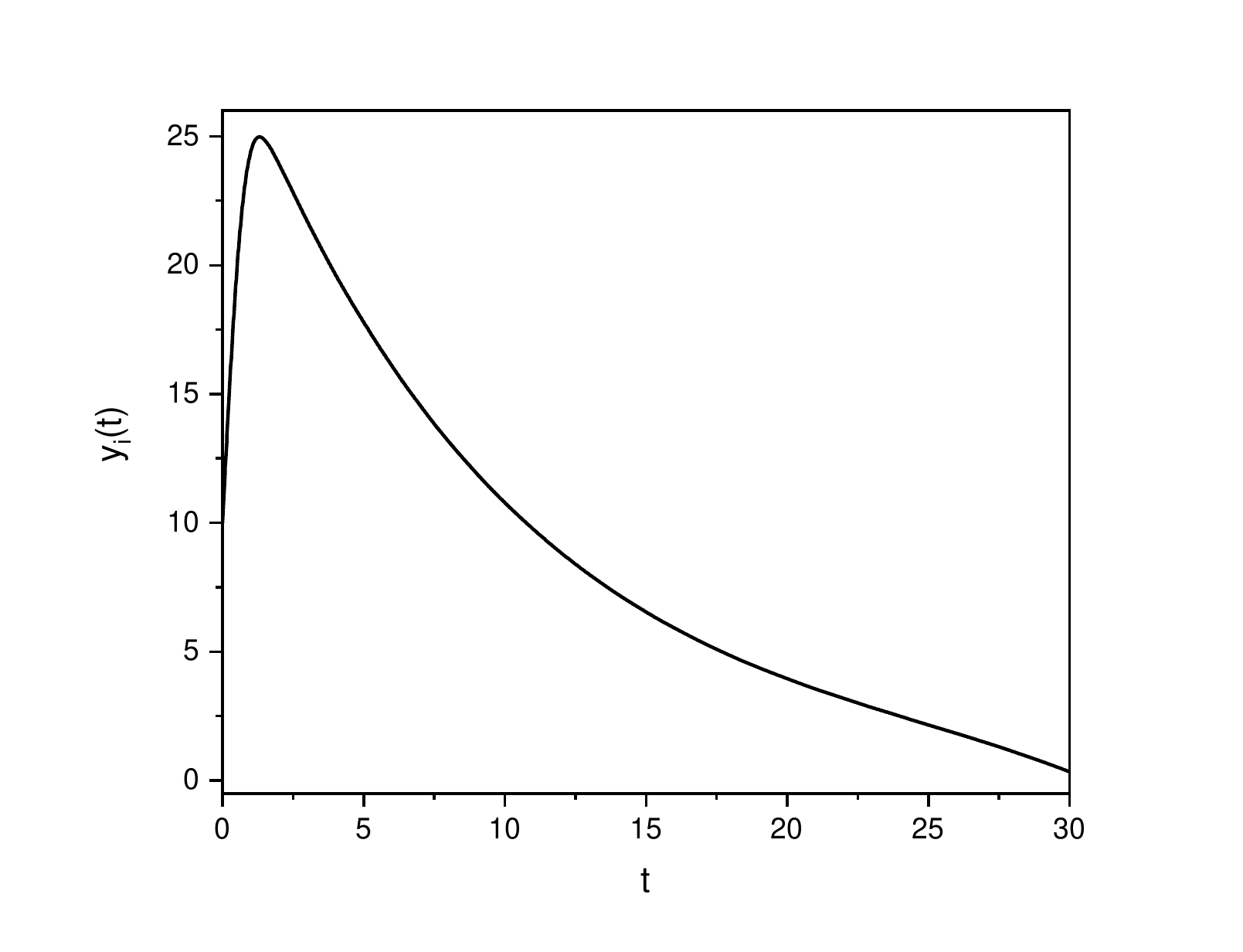}}
		\caption{Figure depicting the ``predator-prey system'' without impulse control}
		\label{figure 8}
	\end{figure}

	The impacts of different system\ref{basic system1} parameters using the ``Takagi-Sugeno fuzzy impulse control model'' are shown in figs. \ref{figure 1} - \ref{figure 8}. 
	
	From figure \ref{figure 1} we observe how the ``prey-predator'' population $(x,y_s,y_i)$ is affected by intra-species competition ($e$) at $r=0.5$, $\lambda=1.5$, $\beta=1.0$, $m=0.25$, $n=0.25$, $d_1=10$, $d_2=10$, $d_3=10$, $a=0.2$, $f_1=0.2$, $f_2=0.1$, $\gamma=0.1$, $B=0.2$. Increase in $e$ decreases population of prey.

	The consequence of of prey max time $d_1$ on ``prey-predator system'' is given in fig. \ref{figure 2} at $r=0.5$, $\lambda=1.5$, $\beta=1.0$, $m=0.25$, $n=0.25$, $d_2=10$, $d_3=10$, $e=0.5$, $a=0.2$, $f_1=0.2$, $f_2=0.1$, $\gamma=0.1$, $B=0.2$. This graph clearly indicates that as $d_1$ increases, so does the prey population.

	In fig. \ref{figure 3} we see that as $d_2$ increases, prey and vulnerable predator population increases at $r=0.5$, $e=0.5$, $\lambda=1.5$, $\beta=1.0$, $m=0.25$, $n=0.25$, $d_1=10$, $d_3=10$, $a=0.2$, $f_1=0.2$, $f_2=0.1$, $\gamma=0.1$, $B=0.2$.
	
	The shift on ``prey-predator system'' with maximum period of an infected predator ($d_3$) is given in fig. \ref{figure 4}  at $r=0.5$, $\lambda=1.5$, $\beta=1.0$, $m=0.25$, $n=0.25$, $d_1=10$, $d_2=10$, $a=0.2$, $f_1=0.2$, $f_2=0.1$, $e=0.5$, $\gamma=0.1$, $B=0.2$. This graphic clearly demonstrate that the population of vulnerable predators increases as $d_3$ increases.
	
	The Impact of Allee effect $\beta$ on ``prey-predator system'' is given in fig. \ref{figure 5} at $r=0.5$, $\lambda=1.5$, $m=0.25$, $n=0.25$, $d_1=10$, $d_2=10$, $d_3=10$, $a=0.2$, $e=0.5$, $f_1=0.2$, $f_2=0.1$, $\gamma=0.1$, $B=0.2$. This graph shows that as we are increasing $\beta$, the population of prey and predator increases.

	In figure \ref{figure 6} we demonstrate the effect of death rate of vulnerable predator ($\lambda$) parameter within fuzzy ``impulsive control'' can be seen in fig. \ref{figure 6} at $r=0.5$, $e=0.5$, $\beta=1.0$, $m=0.25$, $n=0.25$, $d_1=10$, $d_2=10$, $d_3=10$, $a=0.2$, $f_1=0.2$, $f_2=0.1$, $\gamma=0.1$, $B=0.2$. As $\lambda$ increases, population of prey and vulnerable predator decreases.
	
	The shift on ``prey-predator system'' $(x,y_s,y_i)$ by varying transmission rate parameter under fuzzy ``impulsive control'' can be noted in fig. \ref{figure 7} at $r=0.5$, $\lambda=1.5$, $\beta=1.0$, $e=0.5$, $m=0.25$, $n=0.25$, $d_1=10$, $d_2=10$, $d_3=10$, $a=0.2$, $f_1=0.2$, $f_2=0.1$, $\gamma=0.1$. This figure demonstrate that as $B$ increases, population of vulnerable predator increases.	
	
	Finally, the characteristics of the population $(x,y_s,y_i)$ of the 3-species without impulse control are shown in figure. \ref{figure 8} by considering every parameter that was acquired with the ``Takagi-Sugeno fuzzy model'' at $r=0.5$, $e=0.5$, $\lambda=1.5$, $\beta=1.0$, $m=0.25$, $n=0.25$, $d_1=10$, $d_2=10$, $d_3=10$, $a=0.2$, $f_1=0.2$, $f_2=0.1$, $\gamma=0.1$, $B=0.2$, with $x(0)=5, y_s(0)=10, y_i(0)=10$, and $t=10$. The graph clearly illustrates how the populations of predators and prey stabilize.
	
	\section{Conclusion}
	
	The ``Allee effects'' are widely acknowledged as among the most crucial determining variables regarding population change. This approach aligns with the inherent complexity and variability found in many biological systems. In this paper, a ``predator-prey'' model is established with an ``Allee effect'' on both populations and disease in prey. To begin with, a nonlinear ``lotka-Volterra prey-predator model'' with infection in prey based on ``fuzzy impulsive'' control was analyzed. The impulse control model was considered in the context of a fuzzy system based on the ``Takagi-Sugeno model''. Then, the local linear thrust systems are combined to obtain a fully repulsive fuzzy system. Meanwhile, various stability theorems prove the asymptotic stability and exponential stability of impulsive fuzzy systems. Finally, the ``fuzzy impulsive'' control method is illustrated with examples of infected, prey and predator systems with impulsive consequences, and the simulation result shows the potential of the plan. The main results of this study are as follows:
	
	In the current manuscript, we show the following:
	\begin{itemize}
		\item Increase in ``Allee effect'' decreases ``prey-predator'' population.
		\item Prey population increases when intra-species competition between predator and prey decreases.
		\item As $d_1$, $d_2$ increases, population of infected prey and predator decreases.
		\item As $d_3$ decreases, the prey and predator population increases.
		\item Decrease in death rate of predator accelerates the ``prey-predator'' population.
		\item Growth in mortality rate of prey leads to decreases prey population.
		\item Rise in predation coefficient of vulnerable and infected prey directs to decrease in prey population.
		\item Increase in transmission coefficient decreases prey population.
	\end{itemize}
	\section{Acknowledgments} 
	The authors wish to express their appreciation for several excellent suggestions for improvements in this paper made by the referees.

	\bibliography{ref}
	\bibliographystyle{aomplain}
	
\end{document}